\documentstyle[aps,epsfig,lineno,amsmath,hhline,multirow,array]{revtex}

\newcommand{\PO}{\rm l \! P }
\newcommand{\RO}{\rm l \! R }
\newcommand{\xpom}{x_{\PO} }

\newcommand{\as}{\mbox{$\alpha_s~$}}

\newcommand{\pom}{{I\!\!P}}

\begin{document}

\title{A global analysis of inclusive diffractive cross sections at HERA }
\author{C. Royon \thanks{%
 Service de Physique des Particules,
CE-Saclay, F-91191 Gif-sur-Yvette Cedex, France},
L. Schoeffel\thanks{%
 Service de Physique des Particules,
CE-Saclay, F-91191 Gif-sur-Yvette Cedex, France},
S.Sapeta \thanks{%
M. Smoluchowski Institue of Physics
Jagellonian University
Reymonta 4, 30-059 Krakow, Poland
}
R.Peschanski \thanks{%
 Service de Physique Th\'eorique,
CE-Saclay, F-91191 Gif-sur-Yvette Cedex, France},
E.Sauvan \thanks{%
CPPM, IN2P3-CNRS et Universiti\'e de la M\'editerran\'ee, F-13288 Marseille
Cedex 09, France}
}
\maketitle

\begin{abstract}
We describe the most recent data on the diffractive structure functions from the H1 and
ZEUS Collaborations at HERA using 
four models. First, a Pomeron Structure Function (PSF) model, in which the Pomeron is
considered as an object with parton distribution functions. Then, 
 the Bartels Ellis Kowalski W\"usthoff (BEKW) approach is discussed, assuming the simplest perturbative description of the Pomeron
using a two-gluon ladder. 
A third approach, the Bialas Peschanski (BP) model,
based on the dipole formalism is then described.
Finally, we discuss the Golec-Biernat-W\"usthoff (GBW) saturation model which takes into account saturation effects.
The best description of all avaible measurements can be achieved with either the
PSF based model or the BEKW approach. In particular, the BEKW prediction allows
to include the highest $\beta$ measurements, which are dominated by higher twists effects and
provide an efficient and compact parametrisation of the diffractive cross section. The two other models
also give a good description of cross section measurements at small $x$ with a small number of parameters.
The comparison of all predictions allows us to identify interesting differences in the
behaviour of the effective pomeron intercept and in the shape of the longitudinal component
of the diffractive structure functions. In this last part, we present some features that can be
discriminated by new experimental measurements, completing the HERA program.

\end{abstract}

\section{Introduction}

After many years, new measurements of the diffractive structure function
have been made  at the
$ep$ collider HERA, where 27.5~GeV electrons or positrons collide with
820 or 920~GeV protons~\cite{f2d97,f2d97b,zeus,zeusb}. The purpose of this paper
is to analyse all these data under three different theoretical approaches, which
 are described in section II. We also
 study the compatibility of all experimental data, 
by fitting all sets together for each model. In this introduction, we remind briefly the
main features of the  models, used in this paper to understand the diffractive interactions.


In a frame where the incident
proton is very fast, the diffractive reaction can be seen as the
deep inelastic scattering (DIS) of a virtual photon on the proton
target, with a very fast proton in the final state.
We can therefore
expect to probe partons  in a very specific way, under the exchange 
of a color singlet state.
Starting with the pioneering theoretical
work of Ref.\cite{ingelman}, the idea of a point-like structure of the Pomeron
exchange opens the way to the determination of its parton (quark and
gluon) distributions, where the Pomeron point-like structure
can be treated in a similar way as (and compared to) the proton one. Indeed,
leading twist contributions to the proton diffractive structure functions can be
defined by factorisation properties \cite{soper} in much the same way as for
the full proton structure functions themselves. As such, they should obey DGLAP
evolution equations \cite{dglap}, and tehrefore allow for perturbative predictions of
their
$Q^2$ evolution.
On a phenomenological
ground,
these Diffractive Parton Distribution Functions (DPDFs) are the basis of MC simulations like RAPGAP \cite {jung} and they
give a comparison  basis with hard diffraction at the Tevatron~\cite{pap2006}, where the
 factorisation properties are not expected to be valid.
Moreover, the study of DPDFs is also a challenge for
the discussion of different approaches and models where other than leading twist
contributions can be present in hard diffraction.
Indeed,  a strong presumptive evidence exists that higher twist effects may
be quite important in diffractive processes contrary to non-diffractive
ones which do not  require (at least at not too small $Q^2$) such
contributions.  In fact, there are models which incorporate non-negligible
contributions from
higher twist components, especially for relatively small masses  of the
diffractive system. One of our goals  in this paper is to
take into account this peculiarity of diffractive processes.

It is also useful to look at $ep$ scattering in a frame
where the virtual photon moves very fast (for instance in the proton
rest frame, where the $\gamma^*$ has a momentum of up to about 50 TeV at
HERA).  The virtual photon can fluctuate into a quark-antiquark pair,
forming a small color dipole.
Because of its large Lorentz boost, this virtual pair has a lifetime
much longer than a typical strong interaction time. 
Since the interaction between the pair and the proton is
mediated by the strong interaction, diffractive events are possible.
An advantage of studying diffraction in $ep$ collisions is that, for
sufficiently large photon virtuality $Q^2$, the typical transverse
dimensions of the dipole are small compared to the size of a hadron.
The interaction between the quark and the antiquark, as well as the
interaction of the pair with the proton, can be treated perturbatively.
With decreasing $Q^2$ the color dipole becomes larger, and at very low
$Q^2$ these interactions become so strong that a description in terms
of quarks and gluons is no longer justified, and the
diffractive reactions become very similar to those in hadron-hadron
scattering.

The outline of the paper is as follows. In section II, we introduce the different models
used to fit the HERA data that are described in section III. In section IV we start the
analysis of these data with the PSF based approach, leading to the concept of parton
distribution functions. We extract DPDFs under several fit conditions and many details
are discussed in appendix A and B. Then, in sections V to VII,  we follow
with fits of HERA data using dipole models, for which we can consider saturation effects. In particular, the case of the two-gluon exchange model (BEKW) is shown to be quite efficient
to reproduce the data over a large kinematic range including the higher-twist part.
Results are discussed in section VIII. All the approaches covered in this paper give
different properties concerning, e.g., the Pomeron intercept dependences and
the longitudinal diffractive structure function. These aspects are discussed and and outlook is given.

\section{Models}
In this section, we discuss the phenomenology of the different models used to fit
the HERA data.

\subsection{Pomeron Structure Function (PSF) model}

\subsubsection{Reggeon and Pomeron contributions}
\label{qcdfit}

\begin{figure}[tb]
\begin{center}
\includegraphics[width=0.45\textwidth]{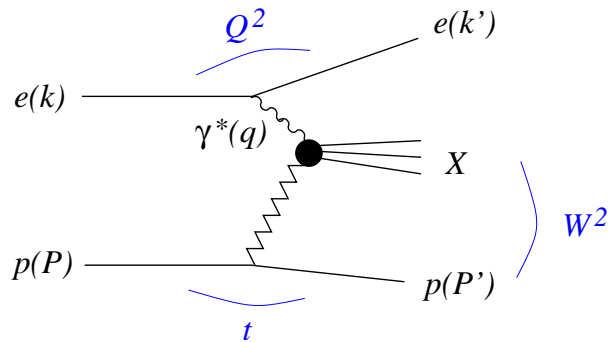}
\caption{Schematic diagram of inclusive diffractive DIS, $ep
  \rightarrow eXp$. Four-momenta are indicated in parentheses.}
\label{diffractive-dis-diagram}
\end{center}
\end{figure}

The diffractive
structure function $F_2^{D(3)}$ 
is investigated in the framework of Regge
phenomenology and expressed as a sum of two factorized 
contributions corresponding to a Pomeron and secondary Reggeon trajectories \cite{pap2006}:

\begin{eqnarray}
F_2^{D(3)}(Q^2,\beta,x_{\PO})=
f_{\PO / p} (x_{\PO}) F_2^{D(\PO)} (Q^2,\beta)
+ f_{\RO / p} (x_{\PO}) F_2^{D(\RO)} (Q^2,\beta) \ ,
\label{reggeform}
\end{eqnarray}
where  $x_{\pom}$ is the
fractional momentum loss of the incident proton and
$\beta$ has the form of a Bjorken variable defined with respect to the
momentum $P-P'$ lost by the initial proton as illustrated on Fig.~\ref{diffractive-dis-diagram}\footnote{
In  a partonic interpretation,
$\beta$ is the momentum fraction of the struck quark with
respect to the exchanged momentum $P-P'$ (the allowed
kinematical range of $\beta$ is between 0 and 1).
}.

In the parametrisation defined by Eq. (~\ref{reggeform}),
$F_2^{D(\PO)}$ can be interpreted as the Pomeron structure function  
and $F_2^{D(\RO)}$ as an effective Reggeon structure function,
with the restriction that it 
takes into account various secondary Regge contributions which can hardly be 
separated.
The Pomeron and Reggeon fluxes are assumed to follow a Regge behaviour with  
linear
trajectories $\alpha_{\PO,\RO}(t)=\alpha_{\PO,\RO}(0)+\alpha^{'}_{\PO,\RO} t$, 
such that

\begin{equation}
f_{{\PO} / p,{\RO} / p} (x_{\PO})= \int^{t_{min}}_{t_{cut}} 
\frac{e^{B_{{\PO},{\RO}}t}}
{x_{\PO}^{2 \alpha_{{\PO},{\RO}}(t) -1}} {\rm d} t ,
\label{flux}
\end{equation}
where $|t_{min}|$ is the minimum kinematically allowed value of $|t|$ and
$t_{cut}=-1$ GeV$^2$ is the limit of the measurement. 
The values of the $t$-slopes parameters can be taken from  Ref. \cite{f2d97} :
$\alpha^{'}_{\PO}=0.06$ GeV$^{-2}$, 
$\alpha^{'}_{\RO}=0.30$ GeV$^{-2}$,
$B_{\PO}=5.5$ GeV$^{-2}$ and $B_{\RO}=1.6$ GeV$^{-2}$. 
Following the procedure described in Ref.~\cite{pap2006}, the Pomeron
intercept $\alpha_{\PO}(0)$ is let as a free parameter in the QCD fit
and $\alpha_{\RO}(0)$ is fixed to $0.50$.

The apparent Regge-factorisation breaking, in other words, the fact that we can not separate the 
$x_{\PO}$ from the $\beta$ and $Q^2$ dependence of the diffractive structure function
is explained in these kinds of models by introducing the Pomeron and Reggeon
 trajectories. As we see in the following, other models such as dipole models
show an intrinsic factorisation breaking and it is of great interest to see if
the secondary trajectories are still needed by the fits.

\subsubsection{Extraction of parton densities in the Pomeron}

We assign parton distribution functions to the Pomeron and to
the Reggeon. A simple
prescription is adopted in which the parton distributions of 
both the Pomeron
and the Reggeon are parametrised in terms of non-perturbative input
distributions at some low scale $Q_0^2$. 
For the structure of the sub-leading Reggeon trajectory,
the pion structure function 
\cite{owens} is assumed
with a free global normalization to be determined by the data. 

For the Pomeron, a quark flavour singlet distribution
($z{ {S}}(z,Q^2)=u+\bar{u}+d+\bar{d}+s+\bar{s}$)
and a gluon distribution ($z{\it {G}}(z,Q^2)$) are parametrised 
at an initial scale
$Q^2_0$, such that

\begin{eqnarray}
z{\it {S}}(z,Q^2=Q_0^2) &=& \left[
A_S z^{B_S}(1-z)^{C_S} (1+D_S z + E_S \sqrt{z}) \right]
\cdot e^{\frac{0.01}{z-1}}  \ , \label{quarka} \\
z{\it {G}}(z,Q^2=Q_0^2) &=& \left[
A_G (1-z)^{C_G}   \right]
\cdot e^{\frac{0.01}{z-1}} \ ,
\label{gluona}
\end{eqnarray}

where $z=x_{i/I\!\!P}$ is the fractional momentum of the Pomeron carried by
the struck parton.  In the following, we present the results in terms of quark
density with the hypothesis that $u=\bar{u}=d=\bar{d}=s=\bar{s}$,
which means that any light quark density is equal to $zS/6$.

Concerning the QCD fit technology, we fix  $\alpha_S(M_Z^2)=0.118$ and
the functions $z{\it{S}}$ and $z{\it{G}}$ of Eq.~(\ref{quarka}) and (\ref{gluona}) 
are evolved to higher $Q^2$ using the
next-to-leading order DGLAP evolution equations
with the code of Ref.~\cite{qcdnum}. 
All fits have also been performed with the code of Ref.~\cite{lolo} as a cross check
which leads to the same results within 2\%.
The contribution to
$F_2^{D(\PO)}(\beta,Q^2)$ from charm quarks is calculated in the fixed
flavour scheme using the photon-gluon fusion prescription given
in Ref.\cite{ghrgrs}. 
The contribution from heavier quarks is neglected. 
No momentum sum rule is imposed because of the theoretical uncertainty in
specifying the normalization of the Pomeron or Reggeon fluxes 
and because it is not clear that such a sum rule is appropriate for
the parton distributions of a virtual exchange. 

Note that diffractive distributions are process-independent
functions.  They appear not only in inclusive diffraction but also in
other processes where diffractive hard-scattering factorisation holds.  
The cross section of such a process can be
evaluated as the convolution of the relevant parton-level
cross section with the DPDFs.
For instance, the cross section
for charm production in diffractive DIS can be calculated at leading order
in $\alpha_s$ from the $\gamma^* g \rightarrow c \bar c$ cross section and
the diffractive gluon distribution.  An analogous statement holds for jet
production in diffractive DIS. Both processes have been analysed at
next-to-leading order in $\alpha_s$.
A natural question to ask is whether  the DPDFs
extracted at HERA can be used to describe hard diffractive processes like the
production of jets, heavy quarks or weak gauge bosons in $p\bar{p}$
collisions at the Tevatron. 
The fraction of diffractive dijet events at CDF is a factor 3 to 10
smaller than would be expected on the basis of the HERA data. The same
type of discrepancy is consistently observed in all hard diffractive
processes in $p\bar{p}$ events, see e.g.~\cite{diehl}.  In
general, while at HERA hard diffraction contributes a fraction of order
10\% to the total cross section, it contributes only to about 1\% at the
Tevatron.
This QCD-factorisation breaking observed in hadron-hadron
scattering can be interpreted as a survival gap probability or a soft color interaction
that need
to be considered in such reactions, and a detailed computation of DPDFs from HERA can help 
in a better understanding of this phenomenon.

\subsection{The Bartels Ellis Kowalski Wusthoff (BEKW) model}
This model assumes the simplest perturbative description of the Pomeron by a
 two-gluon ladder \cite{bartels}. Contrary to the QCD fits described in the previous
section, there is no concept of  DPDFs in this approach.
In Ref.~\cite{bartels}, a parametrisation of the diffractive
structure function in
terms of three main contributions is proposed. The first term describes the diffractive production of a $q \bar{q}$ pair from
a transversely polarised photon, the second one the production of 
a diffractive $q \bar{q} g$ system, and the third one the production of a
$q \bar{q}$ component from a longitudinally polarised photon.
We  note
that the higher Fock states such as $q \bar{q} g g$ are not included in this
model as we  discuss it further in the next section.
More explicitely, we consider the modified form of the two-gluon exchange model, as expressed in
 Ref.~\cite{zeus}, and labeled BEKW fits.

The three different contributions read
\begin{eqnarray}
&~& F_2^{D(3)}(q \bar{q}_{_T})  =  A \left( \frac{x_0}{x_{\PO}}
\right)^{n_2} \beta (1-\beta) \ , \\
&~& F_2^{D(3)}(q \bar{q} g_{_T})  =  B \left( \frac{x_0}{x_{\PO}}
\right)^{n_2}  \alpha_S
\left( \ln [\frac{Q^2}{Q_0^2} +1] \right) (1-\beta)^{\gamma} \ ,  \\
&~& F_2^{D(3)}(q \bar{q}_{_L})  =  C \left( \frac{x_0}{x_{\PO}}
\right)^{n_4}  \left( \frac{Q_0^2}{Q^2+Q_0^2} \right)
\left( \ln [\frac{Q^2}{4 Q_0^2 \beta} +7/4] \right)^2
\beta^3 (1-2 \beta)^{2} \ , 
\end{eqnarray}
where
\begin{eqnarray}
n_{2 (4)}= n^0_{2 (4)} + n^1_{2 (4)}
\ln \left[  \frac{Q^2}{Q_0^2}
+1 \right] \ .
\end{eqnarray}

The ansatz for the $\beta$-dependence is motivated by  general features 
of QCD-parton model calculations: at small $\beta$ (large diffractive masses
$M_X$) the spin 1/2 (quark) exchange in the $q\bar{q}$ production 
leads to a
behavior $\sim \beta$, whereas the spin 1 (gluon) exchange in the
$q \bar{q} g$ term corresponds 
to $\beta^0$. For large $\beta$ (small diffractive masses)  
pertubative QCD leads to 
$1-\beta$ and $(1-\beta)^0$ for the transverse and longitudinal
$q \bar{q} $ terms respectively. For the $q\bar{q}g$
term in $q \bar{q} g $ the situation is slightly more complicated : the exponent
$\gamma$ is left as a free parameter. 
Concerning the $Q^2$ dependence, the first two terms are leading twist
whereas the last one belongs to higher twist.  The $\ln Q^2$-terms follow from
QCD-calculations, and they indicate the beginning of the QCD $Q^2$-evolution.
Finally, the dependence on $x_{\PO}$ cannot be obtained
from perturbative QCD and therefore is left free. 
An additional sub-leading trajectory~\cite{owens} is added to the model,
as for the DGLAP based fit, using the pion structure function to describe the
Reggeon exchange.

The free parameters in the fit are the three
normalisation factors $A$, $B$ and $C$,
the $n^0_2$, $n^0_4$,
$n^1_2$ factors which describe the $x_{\PO}$ dependence of
the cross section and are not predicted by QCD, and the $\gamma$
parameter which describes the $\beta$ dependence of the $q \bar{q} g$ term.
In the above prametrisation, $n_4^1$ is set to zero,
$x_0$ is fixed to $0.1$, $Q_0^2$ is taken to be 1 GeV$^2$
and $\alpha_S$ is taken at the fixed value of 0.25.
The sub-leading trajectory~\cite{owens} is added to the model,
with an overall normalisation $n_{\RO}$ as a further free
parameter, the flux definition being the same as in section~\ref{qcdfit}.

\subsection{The Bialas Peschanski (BP) model}
The dipole model \cite{dipole} provides another approach to describe the Pomeron as a perturbative
object where high Fock states are introduced ($q \bar{q} g g$, $q \bar{q} g g g$...).
In this approach, two components contribute to the diffractive
structure function \cite{robi}. First, an elastic component
corresponds to the elastic interaction of two dipole configurations.
It is expected to be
dominant in the finite $\beta$ region, for small relative masses 
of the diffractive system. 
Secondly, there is an inelastic component where the initial
photon dipole configuration is diffractively dissociated in multi-dipole states
by the target. This process is expected to be important at
small $\beta$ (large masses).

A sub-leading trajectory~\cite{owens} is added to the model,
with an overall normalisation $n_{\RO}$ as a further free
parameter, the flux definition being the same as in in section~\ref{qcdfit}.
The different contributions read
\begin{equation}
F_2^{D(3)} = N^{in}  F_2^{D(3),inel}+N_T^{el} F_T^{D(3),qel}+
N_L^{el} F_L^{D(3),qel}+N_{\RO} F_2^{D(3),Reggeon}
\end{equation}

The transverse and longitudinal elastic components can be written as
\begin{eqnarray}
F_T^{D(el)}=12\frac{N_ce^2\as^4}{\pi}
\xpom^{-2\alpha_\pom +1}\;a^3(\xpom)
\log^3\frac{Q}{2Q_0\sqrt{\beta}}\;e^{-a(\xpom)\log^2\frac{Q}
{2Q_0\sqrt{\beta}}}\\
\times\beta(1\!-\!\beta)\;
\left[{_2F_1}\left(-\frac{1}{2},\frac{3}{2};2;1-\beta\right)
\right]^2\ ,
\label{ftdqef}
\end{eqnarray}
\begin{eqnarray}
F_L^{D(el)}=16\frac{N_ce^2\as^4}{\pi}\;
\xpom^{-2\alpha_\pom+1}\; a^3(\xpom)\log^2\frac{Q}{2Q_0\sqrt{\beta}}
\;e^{-a(\xpom)\log^2\frac{Q}{2Q_0\sqrt{\beta}}}\\
\times\beta^2\;
\left[{_2F_1}\left(-\frac{1}{2},\frac{3}{2};1;1-\beta\right)\right]^2\ ,
\label{fldqef}
\end{eqnarray}
with
\begin{equation}
\alpha_\pom = 1+\Delta_{\pom} \left(\frac{1}{2}\right)=1+\frac{\alpha_s N_c}
{\pi}4\log 2\ ,
\label{defap}
\end{equation}
and
\begin{equation}
a(\xpom)=\frac{\pi}{7\zeta(3)} \alpha_s N_c \log \frac{1}{x} .
\end{equation}
The inelastic component is given by
\begin{equation}
F_{T,L}^{D(in)}=2^9\sqrt{\frac{2}{\pi}} 
H_{T,L}\left(\frac{1}{2}\right)\frac{N_ce^2 \alpha_s^5}{\pi^4}
\;\xpom^{-2\alpha_\pom+1}a^3(\xpom)\frac{Q}{Q_0} 
e^{-\frac{a(\beta)}{2}\log^2\frac{Q}{4Q_0}}
 a^{\frac{1}{2}}(\beta)\beta^{-\Delta_{\pom}}\ .
\label{fdinf}
\end{equation}

The free parameters used in the fit are $\alpha_{\PO}(0)$,  
the
exponents for the Pomeron and the secondary reggeon, the normalisations in
front of the inelastic $N^{in}$, transverse elastic $N_T^{el}$, 
longitudinal elastic
$N_L^{el}$  and the Reggeon component
$N_{\RO}$, as well as the parameter  $Q_0$,
 which is a typically
non-perturbative proton scale.

\subsection{Golec-Biernat-W\"usthoff (GBW) saturation model}
\label{saturation}


The saturation model proposed some time ago by Golec-Biernat and 
W\"usthoff \cite{gbw1,gbw2} was formulated in the color dipole picture. 
In this formalism both the inclusive and diffractive cross sections may 
be calculated.
The diffractive structure function $F^{D(3)}_2$ is the sum of three 
contributions\cite{gbw2} :

\begin{equation}
\label{eq:gbw_f2dsum}
F_2^{D(3)}(Q^2,x_{I\!\!P},\beta)\,=\,F_T^{q\bar{q}}+
F_L^{q\bar{q}}+F_T^{q\bar{q} g}\,,
\end{equation}
The incoming virtual photon $\gamma^*$ (transversely or longitudinally 
polarised) which interacts diffractively with the proton forms either 
the $q\bar{q}$ or $q\bar{q}g$ Fock state. The $F_L^{q\bar{q}}$ and 
$F_T^{q\bar{q}}$ components dominate in the region of large or intermediate 
values of $\beta$ respectively, while the term  $F_T^{q\bar{q}g}$ is most 
important at small $\beta$. The longitudinal component $F_L^{q\bar{q}g}$ 
is the higher twist contribution and therefore may be neglected.

The two terms from Eq.~(\ref{eq:gbw_f2dsum}) related to $q\bar{q}$ dipoles are 
given by the following formulae

\begin{equation}
 x_{I\!\!P}F^{D}_{q\bar{q},L}(Q^{2}, \beta, x_{I\!\!P})=
\frac{3\, Q^{6}}{32 \pi^{4} \beta B_D}\cdot \sum_{f} e_{f}^{2} 
\cdot 2\int_{z_0}^{1/2} dz\, z^{3}(1-z)^{3} \Phi_{0},
\end{equation}

\begin{equation}
x_{I\!\!P}F^{D}_{q\bar{q},T}(Q^{2}, \beta, x_{I\!\!P}) =  
\frac{3\, Q^{4}}{128\pi^{4} \beta B_D} \cdot \sum_{f} e_{f}^{2} \cdot
2\int_{z_0}^{1/2} dz\, z(1-z) 
\left\{ \bar{Q}^2\,[z^{2} + (1-z)^{2}] \Phi_{1} 
       + m_f^{2} \Phi_{0}  \right\},   
\end{equation}
where the sum runs over all active flavours $f$, each with charge $e_f$ and mass $m_f$, and integrate over the light-cone momentum fraction of the virtual photon 
carried by the quark or antiquark $z$. The lower limit of the integration is 
given by $z_0 = (1/2)\left(1 - \sqrt{1 - 4m_f^2/M_X^2}\right)$ where $M_X$ 
denotes  the invariant mass of the diffractive system. The functions 
$\bar{Q}^{2}$  and $\Phi_{0,1}$ are defined as  

\begin{equation}
\bar{Q}^{2} = z(1-z)Q^{2} + m_{f}^{2},
\end{equation}
\begin{equation}
\Phi_{0,1}  \equiv  \left[\, 
\int_{0}^{\infty}r d r K_{0 ,1}(\bar{Q} r)\, \hat{\sigma}(x_{I\!\!P},r)\, 
J_{0 ,1}(kr) \right]^2,
\end{equation}
where $\hat{\sigma}(x_{I\!\!P},r)$ is the dipole cross section specific for 
a given model. In the GBW model it has the form \cite{gbw1}
\begin{equation}
\hat\sigma (x,r)\,=\,\sigma_0\,\left\{
1\,-\,\exp\left(-r^2\, Q^2_{\rm sat}(x)/4  \right) \right\},
\qquad Q^2_{\rm sat}(x) = \left(\frac{x_0}{x}\right)^{\lambda}
\end{equation}
which introduces three parameters : the maximal possible value of the dipole 
cross section $\sigma_0$ and two parameters characterizing the saturation 
scale $Q^2_{\rm sat}(x)$  that is $\lambda$ and $x_0$.

The third component in Eq.~(\ref{eq:gbw_f2dsum}) representing $q\bar{q}g$ 
state has the form

\begin{eqnarray}
\label{eq:gbw_qqg}
\lefteqn{x_{I\!\!P}F^{D}_{q\bar{q}g}(Q^{2}, \beta, x_{I\!\!P}) 
=  \frac{81 \beta\, \alpha_{S} }{512 \pi^{5} B_D} \sum_{f} e_{f}^{2} 
\int_{\beta}^{1}\frac{\mbox{d}z}{(1 - z)^{3}} 
\left[ \left(1- \frac{\beta}{z}\right)^{2} +  \left(\frac{\beta}{z}\right)^{2} \right] } \nonumber \\
 & \times & \int_{0}^{(1-z)Q^{2}}\mbox{d} k_{t}^{2} \ln \left(\frac{(1-z)Q^{2}}{k_{t}^{2}}\right) 
   \left[ \int_{0}^{\infty} u \mbox{d}u \; 
          \sigma_{d}(u / k_{t}, x_{x_{I\!\!P}}) 
           K_{2}\left( \sqrt{\frac{z}{1-z} u^{2}}\right)  J_{2}(u) \right]^{2}.
\nonumber \\
\end{eqnarray} 
This formula was obtained  in \cite{gbw2} using the two-gluon exchange approximation. 
In addition, strong ordering of transverse momenta of the quarks and the gluon 
in the $q\bar{q}g$ dipole was assumed. Notice that (\ref{eq:gbw_qqg}) is 
valid only when we consider massless quarks.

Each contribution is proportional to the inverse of diffractive slope $B_D$ 
for which we take $B_D= 6.0\ {\rm GeV}^{-2}$. The $q\bar{q}g$ component is also 
proportional to the strong coupling $\alpha_s$. Since this is not clear which 
value of $\alpha_s$ should be used we made this coupling a parameter in our 
fits which effectively weights the $q\bar{q}g$ contribution to the diffractive 
structure function.

In this paper we use the version of the GBW model with three light quarks 
only. In addition, we assume these quarks to be massless.


\section{Data sets}
\label{datasets}
In this section, we describe the data used in the following fits.
We use the latest $t$-integrated 
diffractive cross section measurements from H1~\cite{f2d97} (H1RAP)
and ZEUS~\cite{zeus} (ZEUSMX) experiments, derived from the diffractive 
process $e+p \rightarrow e+X+Y$,
where the proton stays either intact or in a low mass state $Y$.
These two data sets are fitted independently and compared.

In case of H1 data~\cite{f2d97}, measurements are presented in terms of the $t$-integrated
reduced cross section $\sigma_r^{D(3)}(\xpom,x,Q^2)$ defined by
the relation
$$
\frac{d^3 \sigma^{ep\rightarrow eXY}}{d\xpom \ dx \ dQ^2} = \frac{4 \pi \alpha^2}{xQ^4}
(1-y+ \frac{y^2}{2}) \sigma_r^{D(3)}(\xpom,x,Q^2)
$$
with
$
\sigma_r^{D(3)} = F_2^{D(3)} -\frac{y^2}{1+(1-y)^2} F_L^{D(3)}
$.
We notice that the relation $\sigma_r^{D(3)} = F_2^{D(3)}$ is a very good approximation
except at large $y$.  

ZEUS data are presented in terms of diffractive
cross sections differential in $M_X$,
$d{\sigma^{\gamma^*p}_{diff}}/{ dM_X}$, and converted to
values for the diffractive structure function values $F_2^{D(3)}$~\cite{zeus}. 

Note that these two data sets represent different methods of measurements,
and therefore different domain in $M_Y$, namely $M_Y < 1.6$~GeV in case of H1
and $M_Y < 2.3$~GeV in case of ZEUS.
In the following, all cross section measurements are corrected to the domain  $M_Y < 1.6$~GeV,
after multiplying ZEUS values by the global factor $0.85$~\cite{f2d97,zeus}.

In addition to the two previous data sets, we use the diffractive
cross sections extracted by H1 and ZEUS experiments
from the process $e+p \rightarrow e+X+p$, where the
leading proton is detected~\cite{f2d97b,zeusb}. This way of tagging diffractive
events is of course very interesting but suffers from a limited detection acceptance,
which gives more restricted samples in terms of kinematic coverage. This is why we use
them only in global fits with all available data sets.
Also, in order to keep the compatibility between all data sets,
cross section measurements from the leading proton samples are
corrected to the domain  $M_Y < 1.6$~GeV, by multiplying them with the global factor 
$1.23$~\cite{f2d97} which has been obtained by comparing the H1 data requiring a
rapidity gap and the data when the proton is tagged in the final state in roman pot
detectors.

To summarize, we give the different data sets used in the fits performed in the
following \footnote{Note that not all models are able to describe the full data sets
since their domain of validity is restricted and these additional cuts will be given
while describing the fits}.

\begin{itemize}
\item $\sigma_{r}^D$ measured by the H1 collaboration using the rapidity gap method
\cite{f2d97} labeled H1RAP, this is the default data set which is not corrected
further;
\item $F_2^D$ measured by the ZEUS collaboration using the ``$M_X$ method"
\cite{zeus} labeled ZEUSMX, data multiplied  by a global factor 0.85. Note that the ZEUS experiment
gives directly the values of the structure functions $F_2^D$, 
i.e. reduced cross sections corrected from the contribution of $F_L^D$,
and we use these values in the following;
\item $\sigma_{r}^D$ measured by the H1 collaboration when a proton is detected in roman
pot detectors \cite{f2d97b} labeled H1TAG, data multiplied by the global factor 1.23;
\item $F_2^D$ measured by the ZEUS collaboration when a proton is detected in roman
pot detectors \cite{zeusb} labeled ZEUSTAG, data multiplied by the global factor 1.23.
\end{itemize}

\section{Results using the PSF model}

Before getting into the results of the QCD fits, let us mention a few technical points
how the fits are performed.
We do not fit the initial scale
$Q^2_0$ as in Ref. \cite{f2d97}, such that we get the best $\chi^2$ value and 
thus the best fit, with a fixed
number of parameters. We have chosen to
increase the number of parameters untill a fit result stable under a variation of the input
scale $Q^2_0$ is obtained. As a result, the $\chi^2$ value and the shapes of the
distributions for quarks and gluons do not depend on the choice of $Q^2_0$.
We have checked that we obtain very similar results as in Ref. \cite{f2d97} while 
doing the QCD fit under the
same conditions as in Ref.~\cite{f2d97} (see appendix A).

As mentioned in the previous section, we  
consider first the H1 data set alone (H1RAP). Further restrictions on the kinematical
plane are requested so that the DGLAP based fits are valid.
The two cuts,  $M_X>2$~GeV and $\beta<0.8$, are necessary to avoid a kinematic range
where higher twists effects are expected to be large, which would spoil the quality
of the determination of the DPDFs.

To allow the fit to be stable for the H1 rapidity gap data alone (H1RAP), 
we have to consider 5 parameters for the 
quarks (see Eq.~(\ref{quarka}))
and 2 for the gluon (see Eq.~(\ref{gluona})). 
With a similar cut scenario as in Ref.~\cite{f2d97},
$Q^2\ge 8.5$~GeV$^2$, $M_X>2$~GeV and $\beta<0.8$, we obtain a $\chi^2=169.3$ for
$Q^2_0=3$~GeV$^2$ and $\chi^2=169.8$ for $Q^2_0=1.75$~GeV$^2$, for 190 data points included
into the QCD fit. Only statistical and uncorrelated systematic uncertainties, added in
quadrature, are considered in this analysis. We have also varied the values of
$Q^2_{min}$, and the results are detailed in Appendix B. Following our
method,  we can produce DPDFs stable within variations of $Q^2_{min}$, which is not the case in
Ref.~\cite{f2d97}. It allows us also to fit all data points with 
$Q^2\ge 4.5$~GeV$^2$, whereas the cut is taken at $8.5$~GeV$^2$ in Ref.~\cite{f2d97}.
In the kinematic domain defined by $Q^2\ge 4.5$~GeV$^2$, $M_X>2$~GeV and $\beta<0.8$,
 we obtain a $\chi^2=250.3$ for 240 data points, using
$Q^2_0=3$~GeV$^2$ as the initial scale for DPDFs definition. 

A parallel analysis is followed for the ZEUS data set~\cite{zeus} (ZEUSMX), with
similar results illustrated on Fig.~\ref{fig3}. 
Note that ZEUS measurements have been multiplied by the global factor $0.85$
as explained in section~\ref{datasets}.
The stability of the QCD fit is
obtained with 4 parameters for the quarks ($E_S=0$ in Eq.~(\ref{quarka}))
and 2 for the gluon. In this case, $\chi^2$ values and shapes are stable within 
variations of $Q_0^2$ and no significant dependence of the distributions is
observed as a function of $Q^2_{min}$. 
In the kinematic domain defined by $Q^2\ge 4.5$~GeV$^2$, $M_X>2$~GeV and $\beta<0.8$, 
we obtain a $\chi^2=100.9$ for 102 points and $Q^2_0=3$~GeV$^2$, with the parameters
given in Table I. For $Q^2_0=1.75$~GeV$^2$, we get $\chi^2=99.9$.
Again for this study, only statistical errors are considered in this analysis.

\begin{center}
\begin{tabular}{|c||c|c|c|}
 \hline
 Parameters & H1RAP & ZEUSMX & All data sets \\
 \hline\hline
$Q^2_{0}$     &  $3$~GeV$^2$    &  $3$~GeV$^2$      &  $3$~GeV$^2$      \\
$Q^2_{min}$     &  $4.5$~GeV$^2$    &  $4.5$~GeV$^2$     &  $4.5$~GeV$^2$     \\
 \hline
 $\alpha_{\PO}$ &  1.120 $\pm$ 0.007  &  1.104 $\pm$ 0.005 &  1.118 $\pm$ 0.005 \\
 \hline
 $A_S$    &   0.28 $\pm$ 0.09     &   0.12 $\pm$ 0.02   &   0.30 $\pm$ 0.15      \\
 $B_S$    &   0.13 $\pm$ 0.08       &   -     &   0.14 $\pm$ 0.11   \\
 $C_S$    &   0.38 $\pm$ 0.08       &   0.50 $\pm$ 0.06  &   0.45 $\pm$ 0.13    \\
 $D_S$    &   6.14 $\pm$ 0.82      &   5.65 $\pm$ 1.24   &   5.72 $\pm$ 0.92    \\
 $E_S$    &  -3.98 $\pm$ 0.22       &   -     &  -3.66 $\pm$ 0.19    \\
 \hline
 $A_G$    &   0.24 $\pm$ 0.06      &   0.74 $\pm$ 0.15    &   0.20 $\pm$ 0.06    \\
 $C_G$    &  -0.76 $\pm$ 0.19       &   3.36 $\pm$ 1.16   &   -0.76 $\pm$ 0.21   \\
 \hline
 $N_{IR}(H1RAP)$ &   5.77 $\pm$ 0.55      &   -   &   6.65 $\pm$ 0.47       \\
$N_{IR}(ZEUSMX)$ &   -      &   -  &   -     \\
 $N_{IR}(H1TAG)$ &   -      &   -    &  5.06 $\pm$ 0.52   \\
 $N_{IR}(ZEUSTAG)$ &   -      &   -    &   4.64 $\pm$ 0.48    \\
 \hline
 $\chi^2$ /(nb data points)&   250.3/240      &   100.9/102    &   377.6/444    \\
 \hline
\end{tabular}
\end{center}
\vskip -0.3cm
\begin{center}
{Table I- Pomeron quark and gluon densities parameters for the three different kinds of
fits (H1RAP alone, ZEUSMX alone, and all four combined data sets). 
No Reggeon contribution is necessary for the ZEUSMX data set, as explained in Ref.~\cite{zeus}.
Also, the parameter $B_S$ is found to be  $0$ for this particular fit.
}
\end{center}

\begin{center}
\begin{tabular}{|c|c|c|}
 \hline
 Data set &   $\chi^2$ (stat. and syst. error) & Nb. of data points \\
 \hline
 H1RAP    & 217.9 & 240        \\
 H1TAG     & 27.5 & 57        \\
 ZEUSMX       & 109.5 & 102        \\
 ZEUSTAG    & 22.7 & 45       \\
\hline
\end{tabular}
\end{center}
\vskip -0.3cm
\begin{center}
{Table II- $\chi^2$ values per data set for the global QCD fit  with statistical and systematics errors added in quadrature (see text).
The fit parameters are found to be very close if only statistical errors are used
during the minimisation procedure.
}
\end{center}

In a second step, we combine the four data sets defined in section~\ref{datasets},
with the normalisation factors for each set defined in this previous section.
To realise this global fit, we use by default the total errors in the definition of the
$\chi^2$. The kinematic domain is again restricted to
$Q^2\ge Q^2_{min}$, $M_X>2$~GeV and $\beta<0.8$, and we study the
stability of the results when varying $Q^2_0$ and $Q^2_{min}$ (see Appendix B).
For $Q^2_{min}=4.5$~GeV$^2$, we get a 
$\chi^2$ value of $377.6$ for 444 data points, with the values for each data sets
given in Table II and parameters listed in Table I
\footnote{
We can also perform the QCD fits using the total error for each data set separately, which gives
a $\chi^2$ value of $198.7$ for 240 data points in the case of H1RAP. When we compare to the global QCD analysis 
(with $\chi^2$ value of $377.6$ for 444 data points), we observe that the $\chi^2/dof$  are compatible, which also justifies our approach to combine all data sets.
}. Note that we assign 
a global Reggeon normalisation parameter 
to each data set
for which this contribution is needed. However, the parameters for the Reggeon and Pomeron
trajectories (defining the flux factors of Eq.~(\ref{flux})) are the same
for all data sets, with $\alpha_{\PO}(0)$ considered as a free parameter in the fit.

\begin{figure}[htbp]
\begin{center}
\includegraphics[totalheight=18cm]{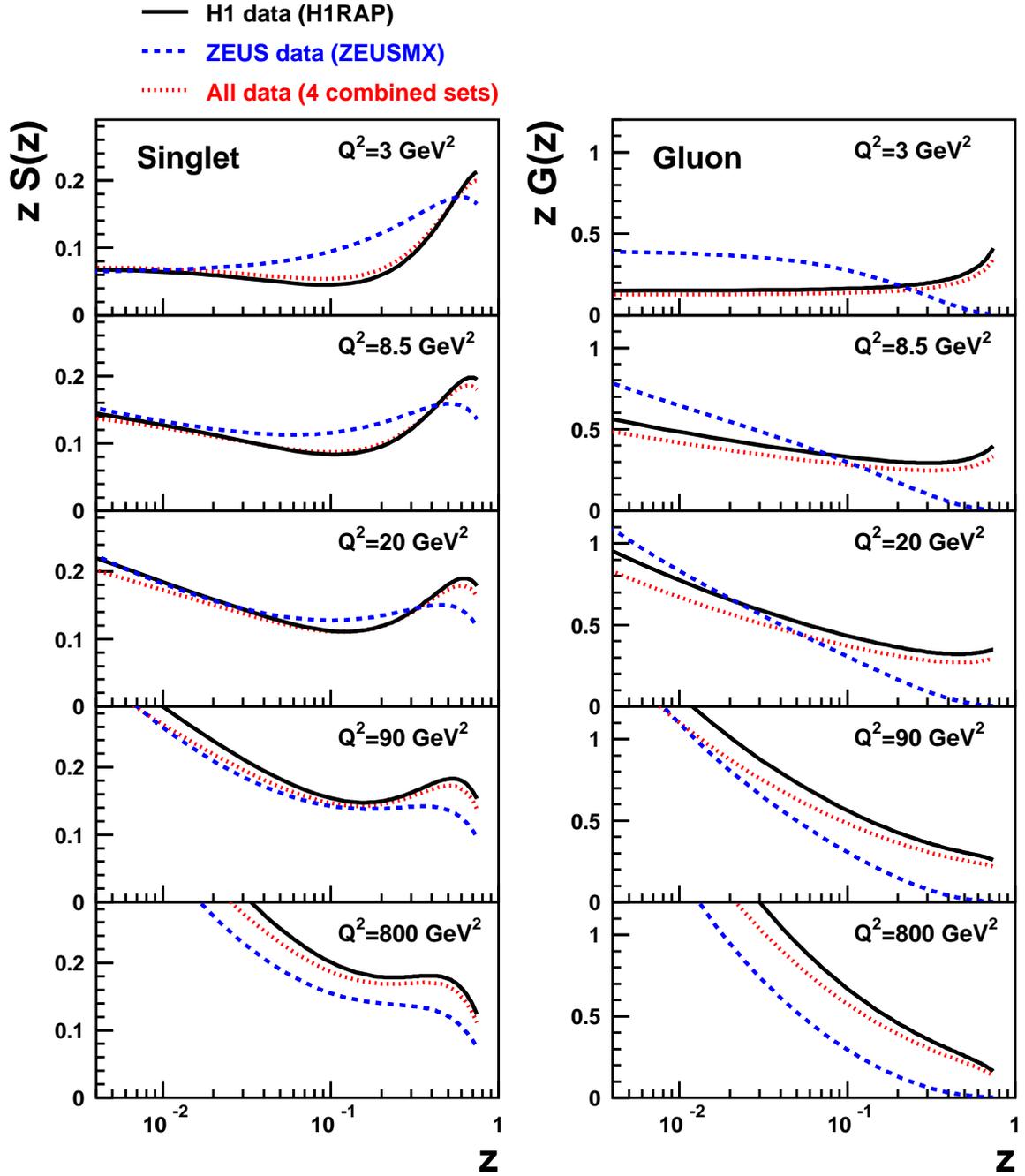}
\end{center}
\caption{ Singlet and gluon distributions 
of the Pomeron as a function of $z$, the fractional momentum of the
Pomeron carried by the struck parton, derived from QCD fits on H1RAP data 
alone, ZEUSMX data alone or the four data sets together. 
The parton densities are normalised to represent 
$\xpom$ times the true parton densities multiplied by the flux factor at
$\xpom = 0.003$. 
}
\label{figfin}
\end{figure}

The fit results are displayed in Fig. \ref{figfin}. We show the quark and gluon
densities in the Pomeron (DPDFs) for H1 data only (H1RAP), ZEUS data only (ZEUSMX) and all four
data combined sets. 
We note that the Pomeron is gluon dominated for all fits. On the
other hand, the shape of the gluon distribution is quite different between the H1 (H1RAP) and
ZEUS (ZEUSMX) data sets, where the high $z$ contribution to the gluon density is small for
the ZEUS data fit. The fact that the parton distributions for the combined fits are
close to the H1 results is simply due to the larger number of H1 data points compared
to ZEUS ones. The gluon density at high $\beta$ is however poorly known
and the uncertainty is of the order of 25 \%. This high $\beta$ region is of 
particular interest for the LHC since it represents for instance a direct background to
the search for exclusive events \cite{pap2006}. More data sets are needed to further constrain this
region such as diffractive jets production at HERA, or Tevatron diffractive jet cross
section measurements.

It is also interesting to show the uncertainty on the high $\beta$ gluon density in the
Pomeron by giving the uncertainty on the parameter $\nu$ if the gluon
density is multiplied by $(1-\beta)^{\nu}$. 
The value of $\nu$ from the fits to the H1RAP and ZEUSMX data sets
respectively is found to be $0 \pm 0.05$ for a variation of $\chi^2$ of $15$ units for the 4 combined 
data sets,
reflecting the large uncertainty on the gluon density at large $z$.


\section{Results using the BEKW model}
\label{bartsec}

Using formulae of section (II.B.1), we can perform fits of the 
structure function measurements (or reduced cross sections) over the whole kinematic
range, including the highest $\beta$ or lowest $M_X$ values, as the longitudinal
component of this model behaves as a higher-twist part.
In a first step, a fit is performed on H1RAP data alone, 
with only one selection cut $Q^2 > 4.5$~GeV$^2$ : we obtain a $\chi^2=286.8$ for
 for 247 data points. Only statistical and uncorrelated systematic uncertainties, added in
quadrature, are considered.
A parallel analysis is performed for the ZEUSMX data set, with
a $\chi^2=193.5$ for
 for 142 data points. Parameters are given in Table III for these two fits.

Then, a global fit is performed to the reduced cross sections of all data sets,
considering for this analysis the total errors (as in section IV). 
The parameters are given in Table IV and
the values of the $\chi^2$ per data sets are presented in Table IV
\footnote{For this model,
we can also perform the BEKW fits using the total error for each data set separately, which gives
a $\chi^2$ value of $226.2$ for 247 data points in the case of H1RAP. 
When we compare to the global BEKW analysis 
(with $\chi^2$ value of $493.1$ for 493 data points), we observe,
as in the previous section, that the $\chi^2/dof$  are compatible.
}.
Fig.~\ref{bekw4} shows the good description of the diffractive
structure function measurements (H1RAP and ZEUSMX) by
the fit result over the whole kinematic range. Only two data sets are presented on this Fig.~\ref{bekw4},
namely~\cite{f2d97,zeus}, to keep the plot lisible. From this Fig.~\ref{bekw4}, we observe that,
taking into account the term $F_2^{D(3)}(q \bar{q}_{_L})$,  the highest $\beta$ bin is properly described
except for the lowest $Q^2$ values. We notice also that the prediction for $\xpom \sigma_{red}^{D}$
is decreasing at low $\xpom$ values for $\beta=0.9$ : it is a reflection of the influence of the
longitudinal component, which becomes large in this domain (see section III). If we apply the cuts
$M_X \ge 2$~GeV and $\beta \le 0.8$ to avoid this kinematic domain, the $\chi^2$ for H1RAP 
improves considerably with a value of 226.6 for
240 data points and for H1TAG, we get a $\chi^2$ of 33.9 for 57 data points.
For the other data sets (ZEUSMX and ZEUSTAG) the $\chi^2/dof$ is stable. The improvement for H1RAP
comes mainly from the kinematic domain $Q^2 \le 15$~GeV$^2$ and $\beta=0.9$, which is not well
described by the fit.

\begin{center}
\begin{tabular}{|c|c|c|c|}
 \hline
 Parameters & H1RAP & ZEUSMX & All data sets   \\
 \hline\hline
 $A(q \bar{q}_{_T})$   &   6.17 \ $10^{-2}$ $\pm$ 0.35  \ $10^{-2}$&   6.28 \ $10^{-2}$ $\pm$ 0.24 \ $10^{-2}$ &   5.98 \ $10^{-2}$ $\pm$ 0.26  \ $10^{-2}$     \\
 $B(q \bar{q} g_{_T})$  &   0.72 \ $10^{-2}$$\pm$ 0.04  \ $10^{-2}$  &   0.70 \ $10^{-2}$$\pm$ 0.03  \ $10^{-2}$ &   0.69\ $10^{-2}$ $\pm$ 0.03  \ $10^{-2}$   \\
 $C(q \bar{q}_{_L})$   &   9.55 \ $10^{-2}$$\pm$ 5.94 \ $10^{-2}$  &   7.59 \ $10^{-2}$$\pm$ 1.15 \ $10^{-2}$ &   6.57 \ $10^{-2}$$\pm$ 1.30  \ $10^{-2}$     \\
 $n_2^0$   &   0.09 $\pm$ 0.01       &   0.18 $\pm$  0.01   &   0.12 $\pm$ 0.01      \\
 $n_2^1$   &  0.04 $\pm$ 0.01   &  0.01 $\pm$ 0.005   &  0.04 $\pm$ 0.004       \\
 $n_4^0$   &   0.35 $\pm$ 0.14  &   0.10 $\pm$ 0.04  &   0.15 $\pm$ 0.43     \\
 $n_4^1$  &  0.   &  0.    &  0.    \\
 $\gamma$   &   9.45 $\pm$ 0.40    &   9.32 $\pm$ 0.54    &   9.76 $\pm$ 0.40     \\
 $x_0$   &  0.1   &  0.1 &  0.1      \\
 $Q_0^2$   &  1.   &  1.  &  1.      \\
\hline
 $N_{IR} (H1RAP)$ &   5.39 $\pm$ 0.57  &   -  &   5.97 $\pm$ 0.49      \\
 $N_{IR} (H1TAG)$  &   -   &   -    &   4.29 $\pm$ 0.53      \\
 $N_{IR} (ZEUSTAG)$ &   -  &   -   &   3.75 $\pm$ 0.50      \\
\hline
 $\chi^2$ /(nb data points)&   286.8/247      &   193.5/142    &   493.1/493    \\
 \hline
\end{tabular}
\end{center}
\vskip -0.3cm
\begin{center}
{Table III- Parameters for the BEKW fit performed in the range 
$Q^2\ge 4.5$~GeV$^2$.
}
\end{center}

\begin{center}
\begin{tabular}{|c|c|c|}
 \hline
 Data set & $\chi^2$ (total error)  & Nb. of data points \\
 \hline
 H1RAP    &   308.3 & 247        \\
 H1TAG     &   44.6 & 59        \\
 ZEUSMX       &   116.5 & 142        \\
 ZEUSTAG   &   23.7 & 45       \\
\hline
\end{tabular}
\end{center}
\vskip -0.3cm
\begin{center}
{Table IV- $\chi^2$ values (total errors) per data set for the global BEKW fit (see text).
}
\end{center}

\begin{figure}[htbp]
\begin{center}
\epsfig{figure=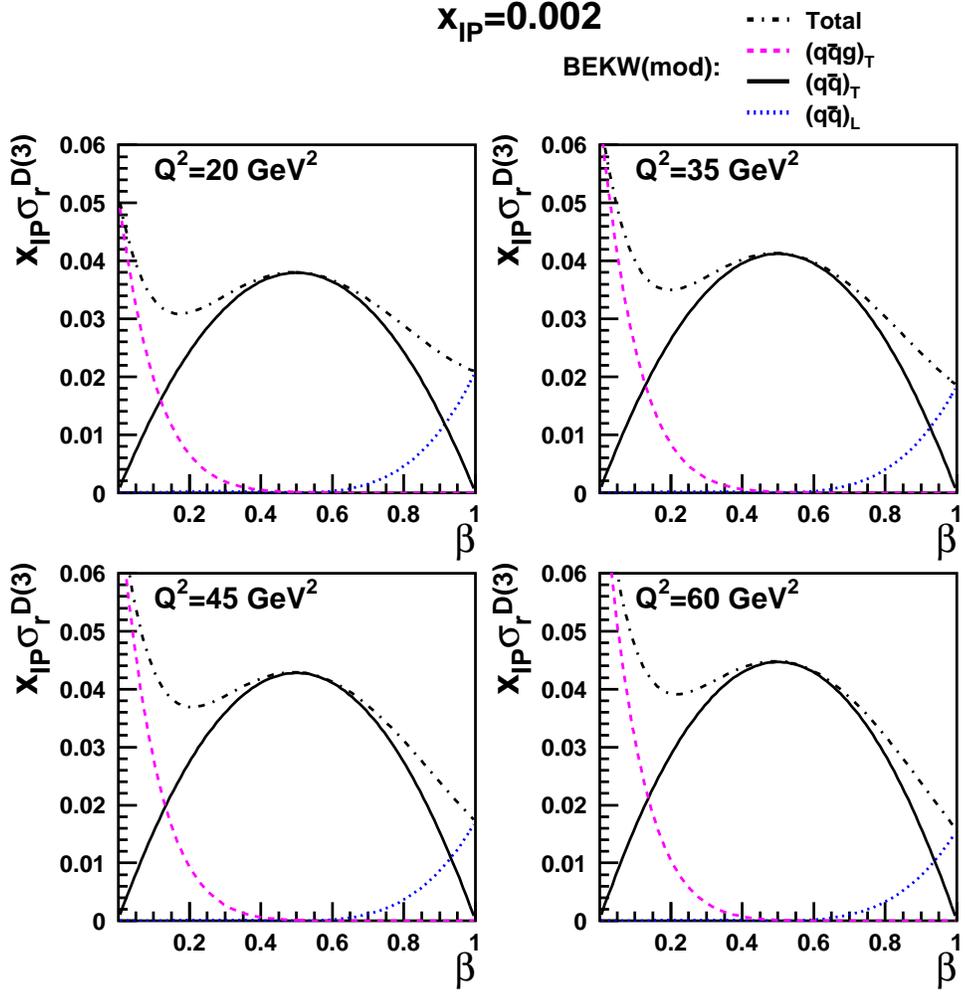,width=15.0cm,
 bbllx=43pt,bblly=174pt,bburx=560pt,bbury=680pt}
\end{center}
\caption{ The three different contributions of the two-gluon exchange model (BEKW), 
fitted on all data sets.
Results are presented as a function of $\beta$, for
different $Q^2$ values, for a fixed $\xpom=0.002$ value.
}
\label{bekw2}
\end{figure}

\begin{figure}[htbp]
\begin{center}
\epsfig{figure=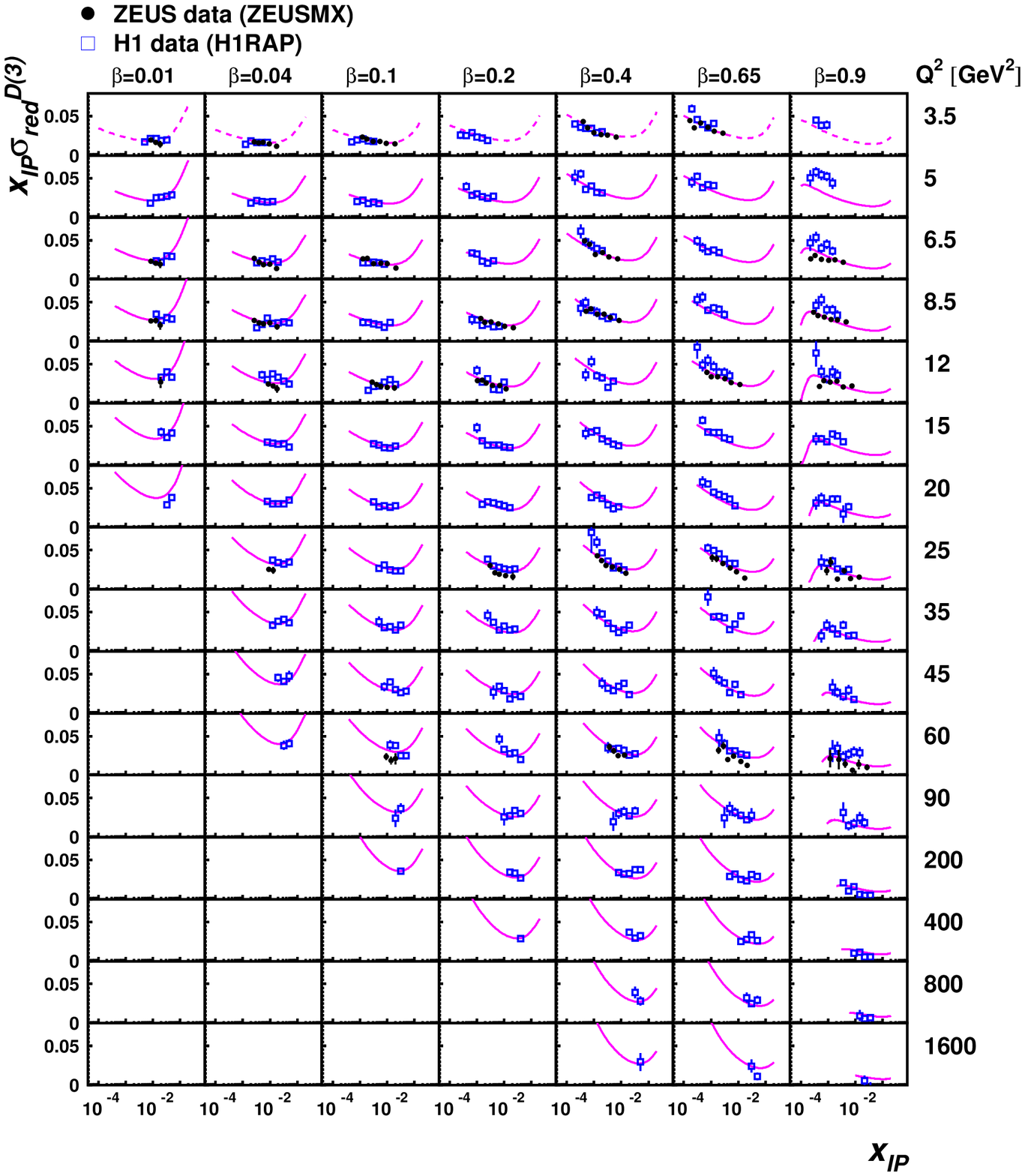,width=18.0cm,
 bbllx=10pt,bblly=70pt,bburx=580pt,bbury=715pt}
\end{center}
\caption{ Comparison of H1 (H1RAP) and ZEUS (ZEUSMX) data sets with the prediction of the BEKW global fit.
Only the statistical part of the uncertainty is shown for the data points on this plot.
A dashed line is drawn for the prediction of the fit on points not included in the analysis.}
\label{bekw4}
\end{figure}

The resulting value of $\gamma$ is large ($\ge 9$), leading to a strong $\beta$
dependence of
the $q \bar{q} g$ term, which is dominant only in the low $\beta$
region. In this kinematic domain,
contributions in which more than two partons emerge
from the diffractive hard scattering have also been found to
be important in measurements of hadronic final state observables in
diffractive DIS \cite{thrust,hfs}. The
longitudinal $q \bar{q}$ term becomes large for $\beta > 0.5$,
indicating a strong higher twist
contribution in this kinematic region. At intermediate $\beta$, the
transverse
$q \bar{q}$ term, varying as $\beta (1-\beta)$ is dominant.
These properties are illustrated in Fig.~\ref{bekw2} for $\xpom=0.002$.
This shows that the scaling
violations of diffractive structure functions observed experimentally
can be described by the $Q^2$ dependence of
the $q \bar{q}$ term at medium $\beta$ since the exponent
of $x_{\PO}$ depends on $Q^2$.

\section{Results using the BP model }
\begin{figure}[htbp]
\begin{center}
\epsfig{figure=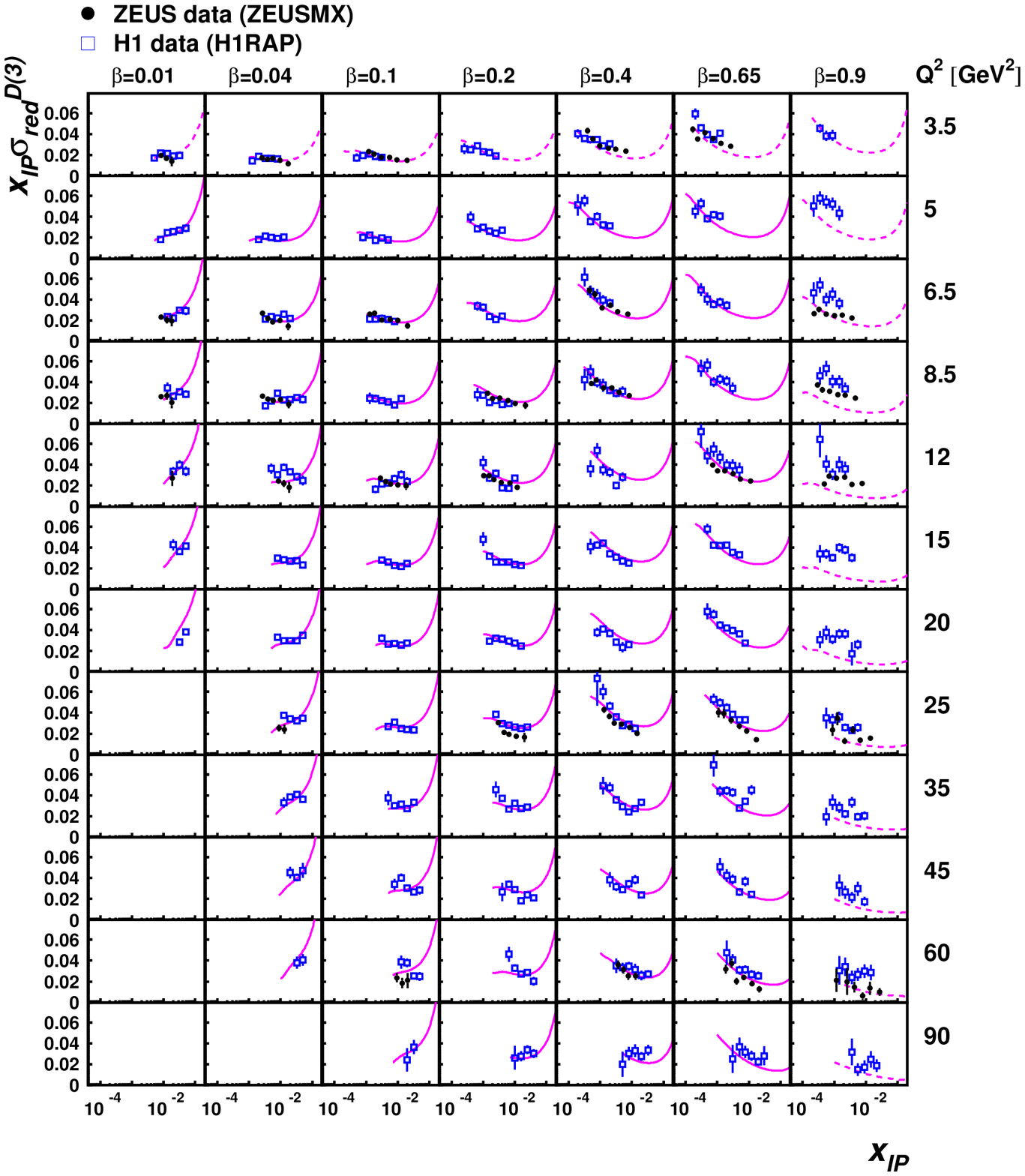,width=18.0cm,
 bbllx=10pt,bblly=70pt,bburx=580pt,bbury=715pt}
\end{center}
\caption{ Comparison of H1 (H1RAP) and ZEUS (ZEUSMX) data sets with the prediction of the BFKL global fit.
Only the statistical part of the uncertainty is shown for the data points on this plot.
A dashed line is drawn for the prediction of the fit on points not included in the analysis.
}
\label{bfkl4}
\end{figure}

In this section, we give the results of the fits based on the dipole
model. Using formulae of section (II.B.2), we can perform fits of the 
structure function measurements in the kinematic domain
$4.5 < Q^2 < 120$~GeV$^2$, $M_X \ge 2$~GeV and $\beta \le 0.8$.
First, a fit is performed on H1RAP data alone,  
we obtain a $\chi^2=324.0$ for
 for 175 data points. Only statistical and uncorrelated systematic uncertainties, added in
quadrature, are considered.
A parallel analysis is followed for the ZEUSMX data set, with
a $\chi^2=85.8$ for 56 data points. 
Then, a global fit is performed to all data sets,
considering the total errors. Parameters are given in Table V and the 
$\chi^2$ values per data set are shown in Table VI.
Fig.~\ref{bfkl4} shows the result of the global fit compared to the data sets H1RAP and ZEUSMX.
In the kinematic domain under study, the agreement is good but the extrapolation in the highest $\beta=0.9$
fails to describe the data.
We note also that in these fits, the Reggeon component is found to be very low, as
the inelatic component is giving a similar behaviour. For the global fit, all
Reggeon normalisations are found to be compatible with zero, but for H1RAP, for which we
get $N_{\RO} = 2.0 \pm 0.5$.  

The value of $\alpha_\pom$
is found to be consistent with the expected intercept for a hard BFKL Pomeron
\cite{bfkl}. This intercept is higher than the value obtained from the
fit to the structure function $F_2$ \cite{dipole}.  
$Q_0$ is very close to the value obtained in the proton structure function
fit. It should be noted that the scale $Q_0$ appears in a non-trivial
way as the virtuality in the inelastic ($Q/4Q_0$) and the
elastic ($Q/2 \sqrt{\beta} Q_0$) components. The validity 
of these results can
be checked by starting with two different values of $Q_0$ 
for each component and
the result of the fit leads to same values within errors, with the same
$\chi^2$. Furthermore, imposing a constant scale for the elastic component, 
 $Q/2 \sqrt{\beta} Q_0 \to Q/Q_0,$
leads to a very bad quality fit. 

\begin{center}
\begin{tabular}{|c|c|c|c|}
 \hline
 Parameters & H1RAP & ZEUSMX & All data sets   \\
 \hline\hline
 $\alpha_{\PO}$    &   1.407   $\pm$ 0.001 &   1.329  $\pm$ 0.006         &   1.339   $\pm$ 0.006    \\
 $Q_0$             &   0.645   $\pm$ 0.005 &   0.266   $\pm$ 0.010         &   0.352   $\pm$ 0.012 \\
 $N^{in}$          &   0.004   $\pm$ 0.0001 &   0.003   $\pm$ 0.001        &   0.003   $\pm$ 0.001      \\
 $N_T^{el}$        &   28.105  $\pm$   1.167   &   154.11  $\pm$   18.873  &   124.230  $\pm$   13.411\\
 $N_L^{el}$        &   1.129   $\pm$  0.045  &   33.904   $\pm$  4.151     &   27.330   $\pm$  2.950  \\
\hline
 $\chi^2$ /(nb data points)&   324./175      &   85.8/56    &   601.3/478    \\
 \hline
\end{tabular}
\end{center}
\vskip -0.3cm
\begin{center}
{Table VII- Parameters for the BFKL fits (see text).
}
\end{center}

\begin{center}
\begin{tabular}{|c|c|c|}
 \hline
 Data set & $\chi^2$ (total error)  & Nb. of data points \\
 \hline
 H1RAP    &   361.6 & 232        \\
 H1TAG     &   32.8 & 59        \\
 ZEUSMX        &   189.5 & 142        \\
 ZEUSTAG  &   17.4 & 45       \\
\hline
\end{tabular}
\end{center}
\vskip -0.3cm
\begin{center}
{Table VIII- $\chi^2$ values (total errors) per data set for the global BFKL fit (see text).
}
\end{center}

\section{Results using the GBW saturation model}

In this section, we give the results of the fits based on the dipole
model including saturation effects. 
Using formulae of section (II.C), we can perform fits of the structure function measurements (or reduced cross sections) over the whole range of $\beta$ and $M_X$ values, as the longitudinal component of this model behaves as a higher-twist part. Since, as in other models, also in this case the sub-leading Reggeon part is added to the three terms in Eq.~(\ref{eq:gbw_f2dsum}) the data for all values of $x_{I\!\!P}$ are used in the fit. 

As shown in \cite{Bartels:2002cj,Golec-Biernat:2006ba} the GBW saturation model works fairly well when fitted to the inclusive $F_2$ data only when $Q^2$ is smaller than $25\ \rm{GeV^2}$. This is because the $Q^2$ evolution which is expected to be important especially at high values of $Q^2$ is not present in the GBW model. This evolution introduced in \cite{Bartels:2002cj} modifies the small $r$ part of the dipole cross section to which the inclusive $F_2$ at high $Q^2$ is very sensitive. However the situation for diffractive structure function $F^D_2$ is different since diffraction is sensitive rather to the intermediate $r$ part of the dipole cross section which is essentially the same regardless of the presence or absence of $Q^2$ evolution. 
In fact, we have checked that imposing the $Q^2 <25\ \rm{GeV^2}$ cut in the fit results in almost no improvement of the fit quality in terms of $\chi^2$ value. Hence, we decided to fit the data using only a $Q^2 > 4.5\ \rm{GeV^2}$ cut as for other models presented in this paper.

\begin{figure}[htbp]
\begin{center}
\epsfig{figure=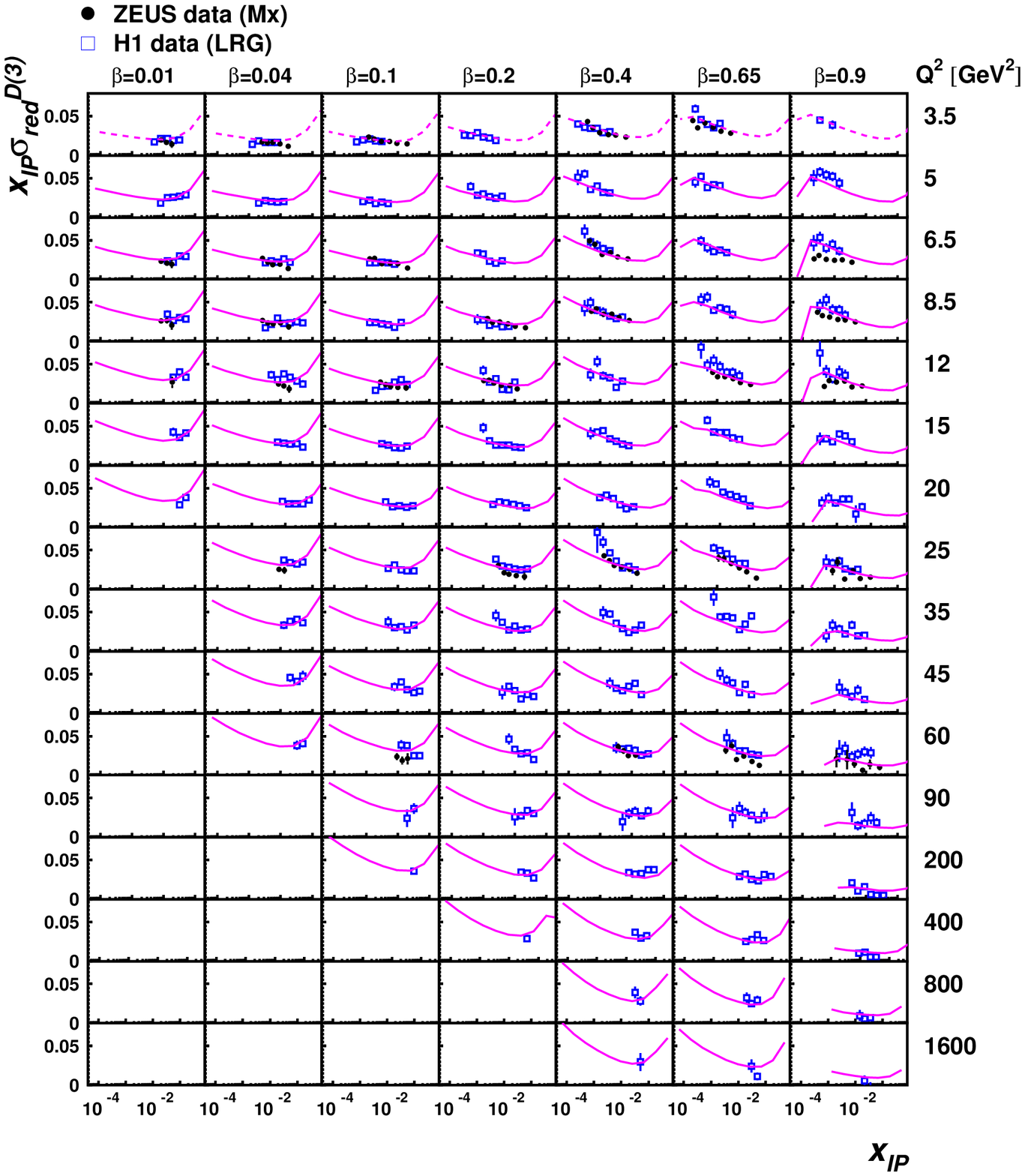,width=18.0cm,
 bbllx=10pt,bblly=70pt,bburx=580pt,bbury=715pt}
\end{center}
\caption{ Comparison of H1 (H1RAP) and ZEUS (ZEUSMX) data sets with the prediction of the GBW global fit.
Only the statistical part of the uncertainty is shown for the data points on this plot.
A dashed line is drawn for the prediction of the fit on points not included in the analysis.}
\label{fig:gbw_red_cs}
\end{figure}

First, a fit is performed on H1RAP data alone,  
we obtain a $\chi^2=272.9$ for 247 data points. 
Only statistical and uncorrelated systematic uncertainties, added in
quadrature, are considered in this analysis.
A parallel analysis is followed for the ZEUSMX data set, with
a $\chi^2=268.0$ for 142 data points. 
Then, a global fit is performed to all data sets and $\chi^2=564.5$ is obtained for 493 data points. Parameters together with $\chi^2$ values per number of data points are given in Table IX. 
Fig.~\ref{fig:gbw_red_cs} shows the result of the combined fit togheter with the H1RAP and ZEUSMX data. A good description is found.

\begin{center}
\begin{tabular}{|c|c|c|c|} \hline 
Parameters         & H1RAP            & ZEUSMX  & All data sets  
\\ \hline  \hline 
$\sigma_0\,$[mb]   & 16.45 $\pm$ 0.24 & 45.83 $\pm$ 0.25    & 27.69 $\pm$ 0.14 
\\ 
$\lambda$          & 0.239 $\pm$ 0.012 & 0.16387 $\pm$ & 0.199 $\pm$
\\ 
$x_0$              & $6.78 \cdot 10^{-3} \pm 0.43 \cdot 10^{-3}$ & 
                     $1.59 \cdot 10^{-8} \pm 0.02 \cdot 10^{-8}$ & 
                     $2.539 \cdot 10^{-5}\pm 0.001 \cdot 10^{-5}$ 
\\ 
$\alpha_s$         & 0.131 $\pm$ 0.004 & 0.075 $\pm$ 0.002 & 0.097 $\pm$ 0.002 
\\ \hline
$N_{IR}(H1RAP)$   &  4.60 $\pm$ 0.50 & - &  5.02 $\pm$ 0.41  \\
$N_{IR}(ZEUSMX)$  &  -               & - &  -                \\
$N_{IR}(H1TAG)$   &  -               & - &  2.96 $\pm$ 0.47  \\
$N_{IR}(ZEUSTAG)$ &  -               & - &  3.46 $\pm$ 0.51  \\ \hline
$\chi^2$/(nb data points) & 272.9/247  & 268.0/142 & 564.5/493 \\ \hline
\end{tabular}
\end{center}
\begin{center}
{Table IX- Parameters for the GBW fits performed in the range 
$Q^2\ge 4.5$~GeV$^2$ (see text).
}
\end{center}

\begin{figure}[htbp]
\begin{center}
\epsfig{figure=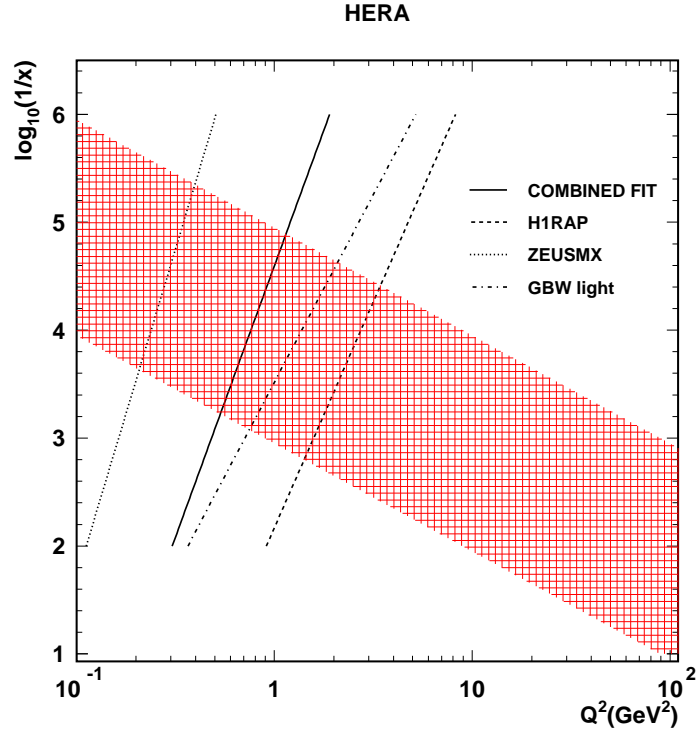,width=10.0cm}
\end{center}
\caption{The critical line in the $(x,Q^2)$-plane obtained with the three sets of parameters given in Table IX. The original GBW result from [23] is shown for reference. The shaded area shows the acceptance region of HERA.}
\label{fig:gbw_crit_line}
\end{figure}

Let us recall that the original GBW parameters from \cite{gbw1} for the fit to $F_2$ data with three light quarks are $\sigma_0 = 23.03\ {\rm mb}$, $\lambda = 0.288$ and $x_0 = 3.04 \cdot 10^{-4}$. The value of $\alpha_s$ used in \cite{gbw2} was fixed at  $\alpha_s = 0.2$. 
Clearly the parameters obtained here for diffractive fits are substantially different from the original GBW parameters. The $\lambda$ parameter is smaller than $0.288$ regardeless of the data set taken to the diffractive fit. The parameter $x_0$ varies a lot depending on which data we use. However, for the combined fit it is one order of magnitude smaller than found in \cite{gbw1}.
This is also reflected in the form of the critical line,
which is an important characteristic of the saturation model. This line in $(x, Q^2)$-plane marks the transition to the domain where the saturation effect are important, \emph{i.e.} to the left of it. 
In Fig.~\ref{fig:gbw_crit_line} we show the critical lines determined with the three sets of parameters from Table IX. Only the parameters which characterize the saturation scale, \emph{i.e.} $\lambda$ and $x_0$, are relevant in this case. The original result from \cite{gbw1} is also presented in Fig.~\ref{fig:gbw_crit_line} for reference. We see that the position of the critical line strongly depends on the data set used in the fit. 
The line obtained with the combined fit parameters is, however, not far away from the GBW result \cite{gbw1} whereas the line corresponding to ZEUSMX parameters is shifted significanly to the left with respect to other lines.


\newpage
\section{Discussion }

In the previous section, we have seen that all models give a correct
description of all data sets. If we compare the $\chi^2/dof$ values per data sets
in the kinematic range $Q^2 > 4.5$~GeV$^2$, $M_X \ge 2$~GeV and $\beta \le 0.8$,
PSF and BEKW fits are compatible, with a global $\chi^2/dof$ of 0.83 and 0.81 (respectively).
When we consider 
in addition the kinematic cut $Q^2 < 120$~GeV$^2$, the BP and GBW approaches lead to  a global $\chi^2/dof$
 of 1.24 and 1.31 (respectively).
Then, PSF and BEKW predictions give a better description of the inclusive diffractive measurements
but the BP and GBW models are also reasonnable and it makes sense to compare some specific features 
of all models considered above.

Already, some properties have been mentionned in section (IV). 
For the BEKW  parametrisation, we have seen that
taking into account the term $F_2^{D(3)}(q \bar{q}_{_L})$ makes it possible
to describe the highest $\beta$ bin ($\beta=0.9$),
except for the lowest $Q^2$ values. Then, this model provides a very efficient 
description of all the inclusive diffractive measurements, with a simple functional
form and few parameters. Of course, this is a leading order approach, as the dipole
and saturation models we have considered in previous sections. The PSF model is
the only one at next-to-leading order, with the other advantage that
DPDFs can be used to predict cross sections in hadron-hadron scattering,
taking into account also the survival gap probability (see section II.A). 

\begin{figure}[htbp]
\begin{center}
 \includegraphics[totalheight=9cm]{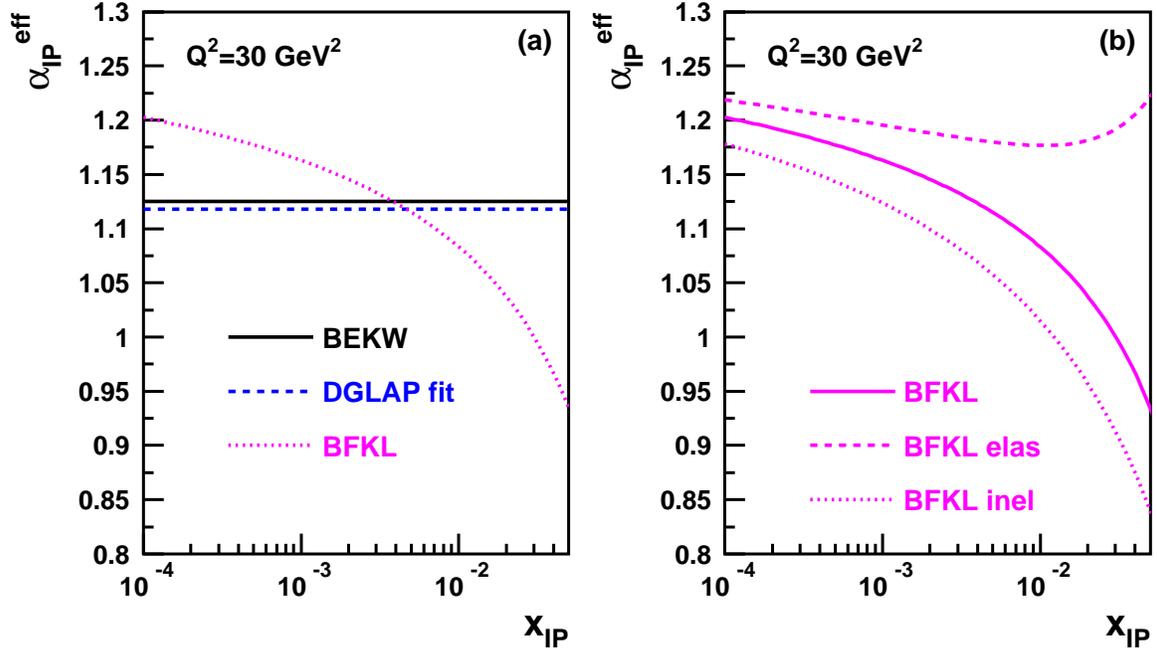}
\caption{ Effective Pomeron intercept ($\alpha_{\pom}=0.5(\frac{d \ln F_2^{D(\pom)}}{d \ln 1/\xpom}+1)$)
as a function of $\xpom$.
The dependence of $\alpha_{\pom}$ on $\xpom$ is shown for all models 
at $Q^2=30$~GeV$^2$ and $\beta=0.3$ : in plot (a) we present predictions for BEKW as a full line, 
PSF fit as a dashed line, BP as a dotted line and GBW as a dashed-dotted line.
A similar plot (b) is shown for the BP model alone, with elastic and inelastic
component drawn separately (dashed and dotted curves respectively).}
\label{Figdisc1}
\end{center}
\end{figure}
\begin{figure}[htbp]
\begin{center}
 \includegraphics[totalheight=9cm]{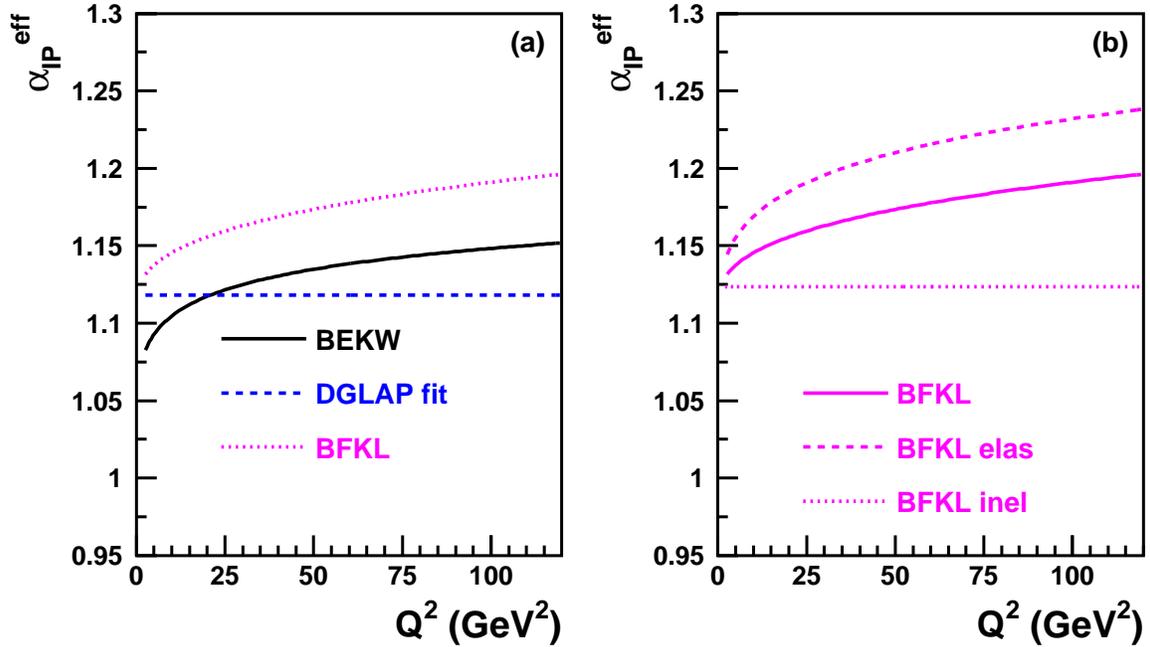}
\caption{ Effective Pomeron intercept as a function of $Q^2$.
The dependence of $\alpha_{\pom}$ on $Q^2$ is shown for all models 
at $\xpom=10^{-3}$ and $\beta=0.3$ : in plot (a) we present predictions for BEKW as a full line, 
PSF fit as a dashed line, BP as a dotted line and GBW as a dashed-dotted line.
A similar plot (b) is shown  for the BP model alone, with elastic and inelastic
component drawn separately.}
\label{Figdisc1b}
\end{center}
\end{figure}

Let us now study the dependence on $\xpom$ which is directly related 
to the rapidity gap dependence. One can define an effective Pomeron 
intercept in the following way:
\begin{eqnarray}
\alpha_{\pom}^{eff} = \frac{1}{2} \left( \frac{d ln F_2^{D}}{d ln 1/ x_{\pom}}
+ 1 \right)
\end{eqnarray}
where the $t$ dependence is integrated out. This effective exponent can also be determinated for the
inelastic and elastic components separately for the BP model.
Results are presented for $\alpha_{\pom}^{eff}$
as a function of
$\xpom$ (Fig.~\ref{Figdisc1}) or $Q^2$ (Fig.~\ref{Figdisc1b}). 
In Fig.~\ref{Figdisc1} (b), the effective intercepts of both
components from the BP model and their sums have been presented. 
The range of obtained values sits essentially 
between these two limits except in the large $\xpom$ region ($\xpom \ge 
10^{-2}$). It is clearly not consistent with the soft Pomeron value
(1.08). The shape observed  on Fig.~\ref{Figdisc1} (b) can be explained by the large logarithmic
corrections induced by the $a^3(x_\pom)$ term, proportional to
$\log^3(1/x_\pom)$, present in both diffractive components (see formulae 
\ref{ftdqef}, \ref{fldqef}, \ref{fdinf}). The effect of this logarithmic
term induces also an $\xpom$ dependence of the intercept. Moreover,
it can be seen that the $\xpom$ dependence of the intercept is different 
between the elastic and the inelastic components. This induces an apparent breaking of
factorisation directly for the diffractive components of this model, which 
comes in addition to the known factorisation breaking due to secondary 
trajectories~\cite{robi}. 
Fig.~\ref{Figdisc1b} shows the $Q^2$ dependence of the Pomeron intercept $\alpha_{\pom}^{eff}$
predicted by all models,
with clear differences between the PSF fit and the BEKW model. In the case of PSF model, the $Q^2$
behaviour is by definition included in the DPDFs and not in the factorised $\xpom$ part. In case of BEKW,
there is a natural $Q^2$ dependence arising in the effective Pomeron intercept. A similar shape is
observed in the case of the BFKL model, with still a normalisation factor compared to BEKW as discussed above.
Note that the significant difference in behaviour between the PSF and the BEKW models is a challenge
for experimentalists and the determination of the $Q^2$ dependence of the Pomeron intercept
would provide a good discrimination between models.

\begin{figure}[htbp]
\begin{center}
 \includegraphics[totalheight=9cm]{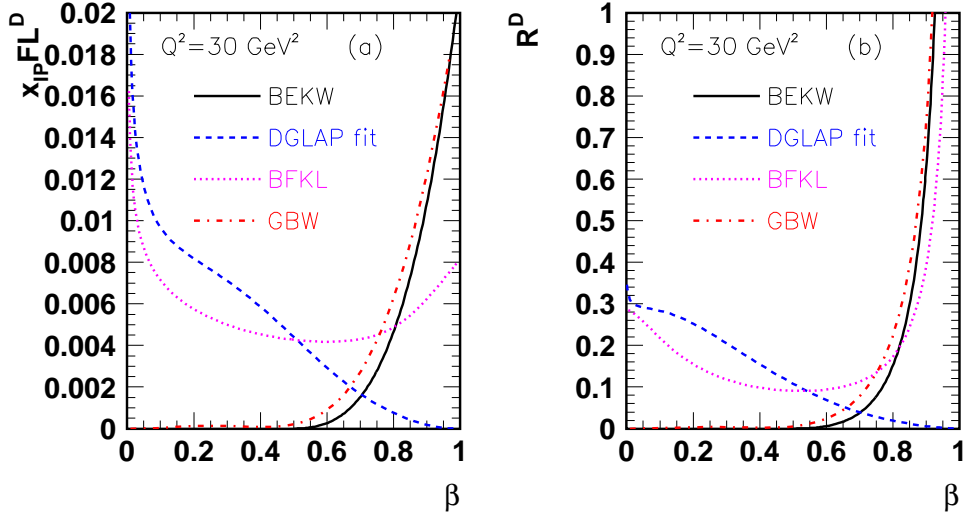}
\caption{ Predictions for $\xpom F_L^{D}$  and $R^D=\frac{F_L^D}{F_2^D-F_L^D}$  as a function of $\beta$
at $Q^2=30$~GeV$^2$ and $\xpom=10^{-3}$. We present predictions for BEKW as a full line, 
PSF fit as a dashed line, BP as a dotted line and GBW as a dashed-dotted line.}
\label{Figdisc2}
\end{center}
\end{figure}

Another important conclusion from the previous analysis is the prediction for the
longitudinal diffractive structure function by all models.
Fig.~\ref{Figdisc2} displays as a function of $\beta$ 
the longitudinal component of the
proton diffractive structure functions (Fig.~\ref{Figdisc2} (a)) and 
the ratio $R$ of the longitudinal to the transverse components (Fig.~\ref{Figdisc2} (b)). 
Here again, we observe significant differences.
The BEKW of GBW models predicts only a longitudinal contribution at large $\beta$, which behaves in $Q^2$ as a higher twist part
(see section II). On the contrary, the PSF model predicts the leading twist part of $F_L^D$ which is increasing at small values
of $\beta$ and tends to $0$ when $\beta$ tends towards $1$. It is interesting to observe that the BP model contains these
two features for the behaviour of $F_L^D$ with a leading twist behaviour at low $\beta$ and a non-zero component for the
largest $\beta$ values.

\begin{figure}[htbp]
\begin{center}
 \includegraphics[totalheight=18cm]{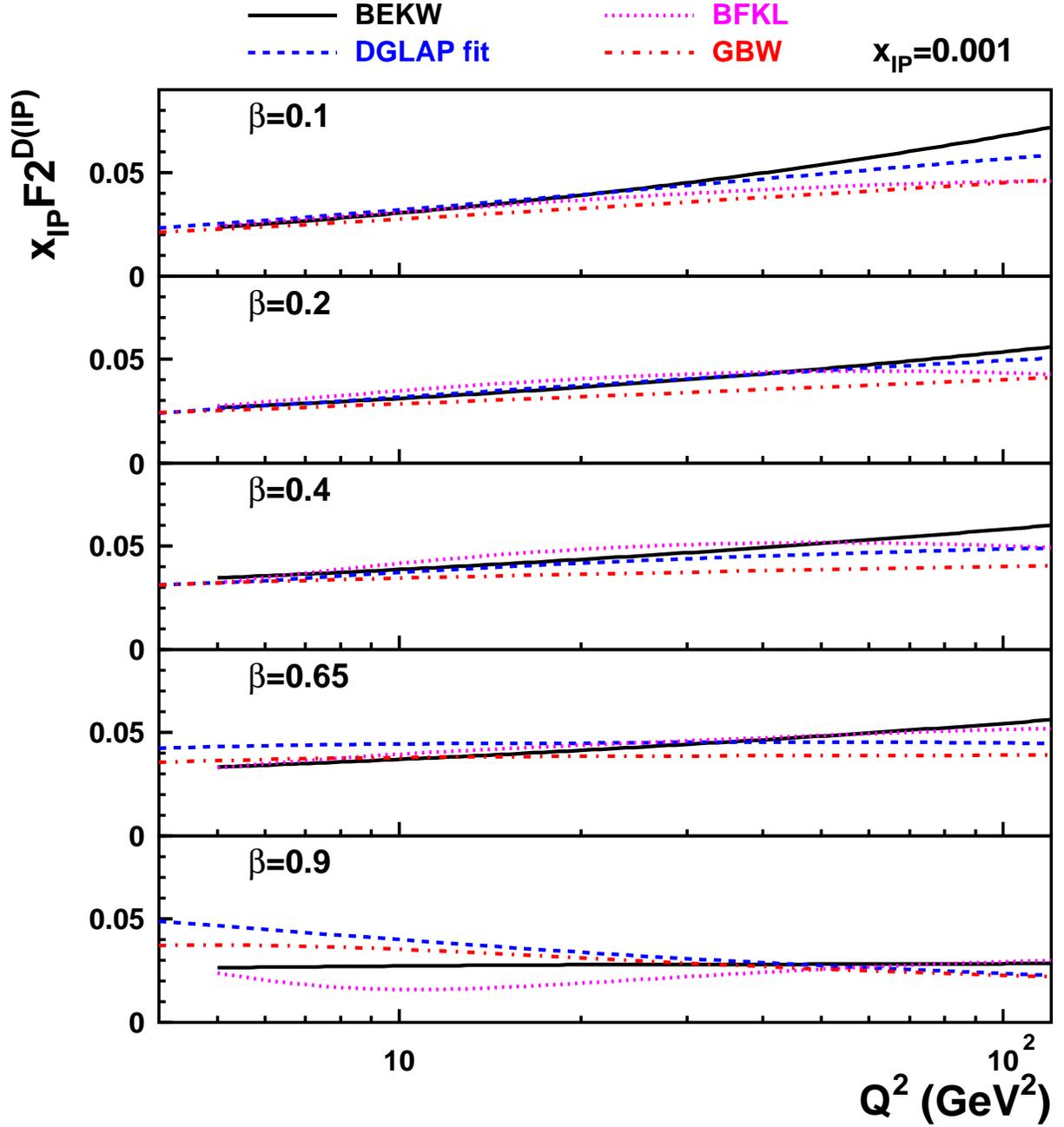}
\caption{ Scaling violations, $\xpom F_2^{D(\pom)}$ as a function of $Q^2$ at different
values of $\beta$ and $\xpom=10^{-3}$. We present predictions for BEKW as a full line, 
PSF fit as a dashed line, BP as a dotted line and GBW as a dashed-dotted line.}
\label{Figdisc4}
\end{center}
\end{figure}

\begin{figure}[htbp]
\begin{center}
 \includegraphics[totalheight=18cm]{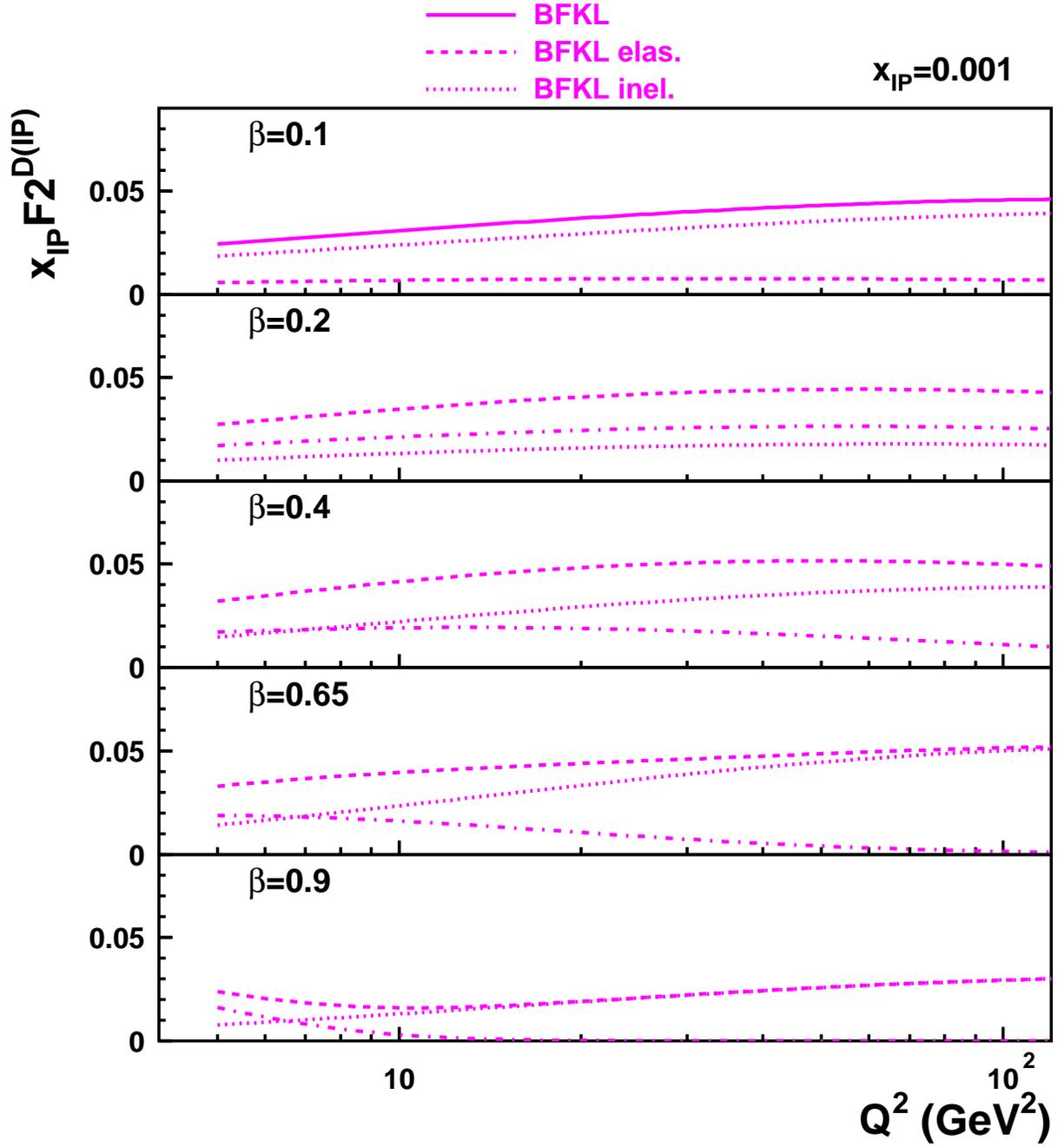}
\caption{ Scaling violations, $\xpom F_2^{D(\pom)}$ as a function of $Q^2$ at different
values of $\beta$ and $\xpom=10^{-3}$. We present the BP model predictions, with
elastic and inelastic components shown separately (dashed and dotted curves respectively).}
\label{Figdisc4b}
\end{center}
\end{figure}

Also, a 
 striking feature of the diffractive proton structure functions concerns the scaling violation, namely
the $Q^2$ dependence at fixed $x_{\pom}$ as a function of $\beta$.
The structure functions are increasing with $Q^2$ even at
very high $\beta$ (see Fig.~\ref{Figdisc4} and Fig.~\ref{Figdisc4b}) at variance with the behaviour of the total proton
structure function as a function of $x$. Note that we have only plotted the Pomeron part of the structure function, then
subtracting the Reggeon component from the total structure function prediction.
We observe that all models are in good agreement with some small differences in the highest $\beta$ bin, but
much smaller than the experimental uncertainty accessible in this domain.

\section{Conclusion and outlook}

We have discussed the most recent data on the diffractive structure functions from the H1 and
ZEUS Collaborations at HERA using 
four models, based on different 
approaches of the vacuum exchange structure. We have shown that
the best description of all avaible measurements can be achieved with either the
PSF  model or the BEKW model. 
In the case of the PSF approach, we have derived the DPDFs of the Pomeron under several variations in
the analysis. In particular, a global analysis of all avaible data has been performed with a proper 
description of these measurements. The main features of the DPDFs, already observed in Ref.~\cite{pap2006},
are conserved. A large gluon content is still present with a large uncertainty of about 25\% at large $\beta$.
Also, we have shown that the BEKW prediction allows
to include  the highest $\beta$ measurements in the analysis. For these values of $\beta$,
 higher twists effects are important and are correctly taken into account by the BEKW model through 
 the longitudinal component of the diffractive cross section. We can mention that this model
provides an efficient and compact parametrisation of the diffractive cross section
over a large kinematic domain.
The BP and GBW models
also give a good description of all cross section measurements at small $x$ with a small number of parameters.
During the discussion section, we have noticed that
 the significant difference in behaviour for the effective Pomeron intercept
 between the PSF and the BEKW approaches is a challenge
for experimentalists and the determination of the $Q^2$ dependence of the $\alpha_{\pom}$
would provide a good discrimination between models.
Concerning the longitudinal diffractive structure function, we have also observed interesting differences between models.
 In particular, the BP approach combines the low and high $\beta$ behaviours of the PSF and
BEKW models respectively.
Once this observable could be accessed
experimentally, it would be of great help for further developpements of the theory.

\section{Acknowledgments}
We would like to thank E. Perez for fruitful discussions on many aspects of DPDFs.
S.S. acknowledges grants of the Polish Ministry of Education and
Science: No. 1 P03B 028 28 (2005-08) and N202 048 31/2647 (2006-08) and
the French--Polish scientific agreement Polonium.

\newpage

\newpage

\newpage

\newpage

\section{Appendix A : Reproducing H1 Fit A }

In Ref.~\cite{f2d97}, the normalisations of the flux of Pomeron and Reggeon 
are defined in a different way, compared to our definitions in this paper or
in our previous analysis (see Ref.~\cite{pap2006}). 
In this appendix, to compare results from our fit  to those of Ref.~\cite{f2d97},
we follow  the same conventions as in Ref.~\cite{f2d97} for the normalisations
of the flux factors. Also, we use the Reggeon contribution from Ref.~\cite{owens}
and we perform a QCD fit under the same conditions as fit A of Ref.~\cite{f2d97} : results
are presented on Table A and  are very close.
Note also that, in our procedure, we use only the statistical and uncorrelated
uncertainties added in quadrature, which gives a $\chi^2$ of 169.9 for 190 data points,
whereas, in Ref.~\cite{f2d97}, the $\chi^2$ definition also includes  parameters
for all correlated systematics.

\begin{center}
\begin{tabular}{|c|c|c|}
 \hline
 parameters & Our fit on H1 data & Table 3 of Ref.~\cite{f2d97} (fit A)  \\
 \hline\hline
$Q^2_{0}$     &  $1.75$~GeV$^2$   &  $1.75$~GeV$^2$        \\
$Q^2_{min}$     &  $8.5$~GeV$^2$   &  $8.5$~GeV$^2$        \\
 \hline
 $\alpha_{\PO}$ &  1.118 $\pm$ 0.008  &  1.118 $\pm$ 0.008  \\
 \hline
 $A_S$    &   1.10 $\pm$ 0.32   &   1.06 $\pm$ 0.32        \\
 $B_S$    &   2.33 $\pm$ 0.35   &   2.30 $\pm$ 0.36       \\
 $C_S$    &   0.61 $\pm$ 0.15   &   0.57 $\pm$ 0.15     \\
 \hline
 $A_G$    &   0.13 $\pm$ 0.03    &   0.15 $\pm$ 0.03    \\
 $C_G$    &  -0.92 $\pm$ 0.16    &  -0.95 $\pm$ 0.20     \\
 \hline
 $N_{IR}$ &   2.8 \ 10$^{-3}$  $\pm$ 0.4 \ 10$^{-3}$     
 &   1.7 \ 10$^{-3}$  $\pm$ 0.4 \ 10$^{-3}$  \\
\hline
\end{tabular}
\end{center}
\vskip -0.3cm
\begin{center}
{Table A- DPDFs parameters (see text).
}
\end{center}

\newpage

\section{Appendix B : The DGLAP fits}

\begin{figure}[htbp]
\begin{center}
\includegraphics[totalheight=18cm]{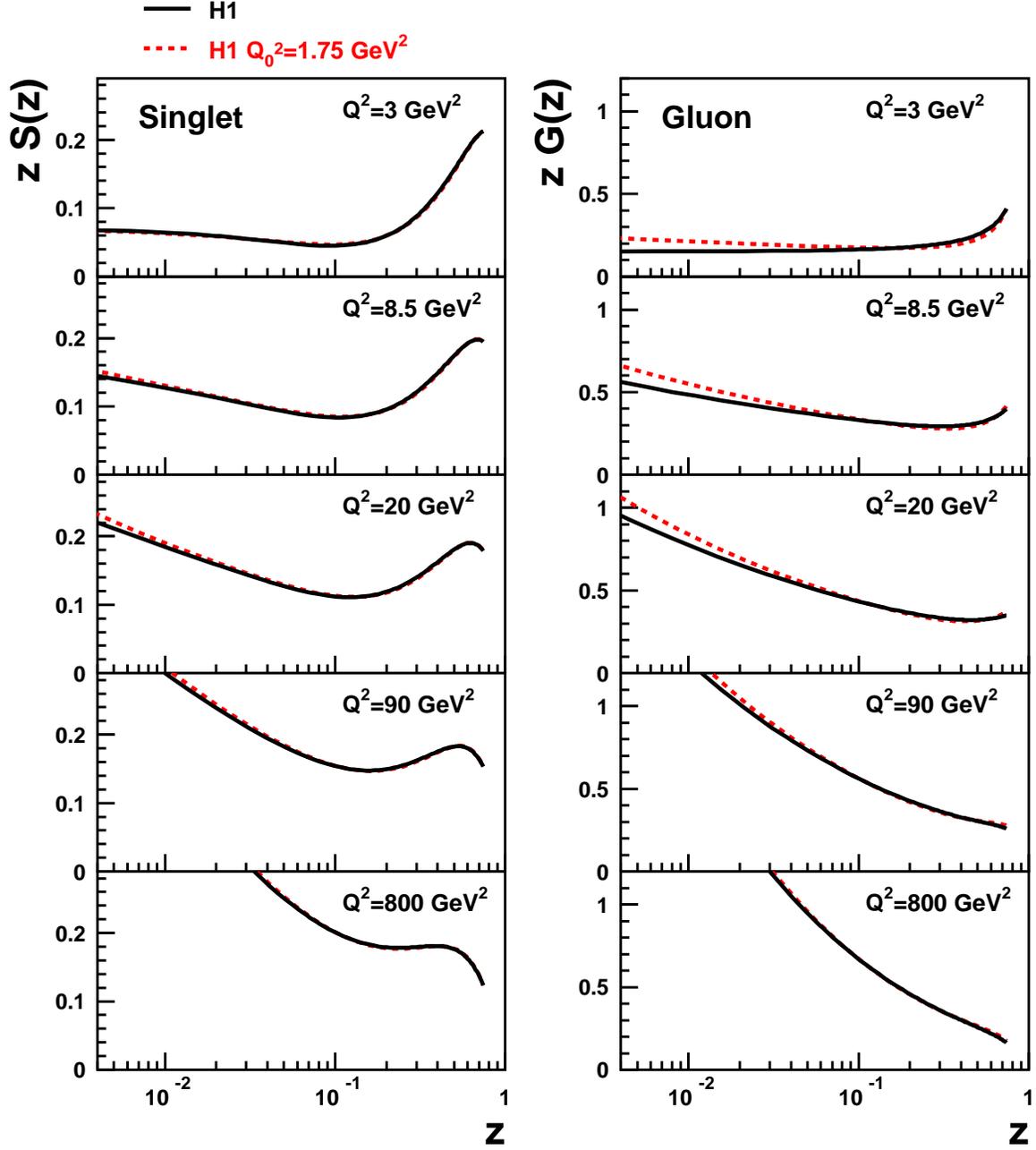}
\end{center}
\caption{ Singlet and gluon distributions 
of the Pomeron as a function of $z$, the fractional momentum of the
Pomeron carried by the struck parton, derived from QCD fits on H1 data. 
Results are presented with $Q^2_0=3$~GeV$^2$ (full lines) and 
$Q^2_0=1.75$~GeV$^2$ (dashed lines). 
The parton densities are normalised to represent 
$\xpom$ times the true parton densities multiplied by the flux factor at
$\xpom = 0.003$. 
}
\label{fig0}
\end{figure}

\begin{figure}[htbp]
\begin{center}
\includegraphics[totalheight=18cm]{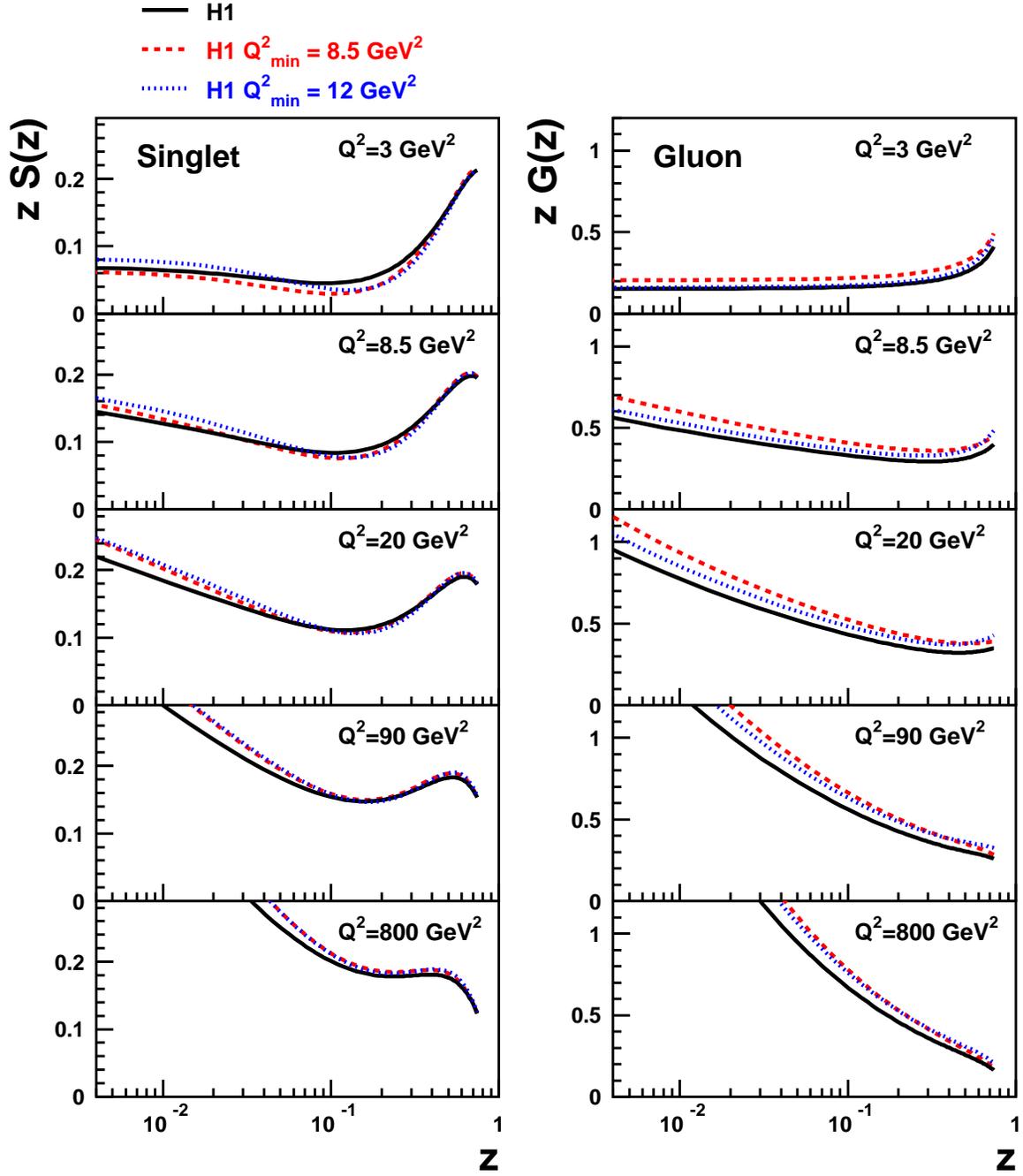}
\end{center}
\caption{ Singlet and gluon distributions 
of the Pomeron as a function of $z$ derived from  QCD fits on H1 data. 
Results are presented with $Q^2_{min}=4.5$~GeV$^2$ (full lines),
$Q^2_{min}=8.5$~GeV$^2$ (dashed lines) and $Q^2_{min}=12$~GeV$^2$ (dotted lines). 
}
\label{fig1}
\end{figure}

\begin{figure}[htbp]
\begin{center}
\includegraphics[totalheight=18cm]{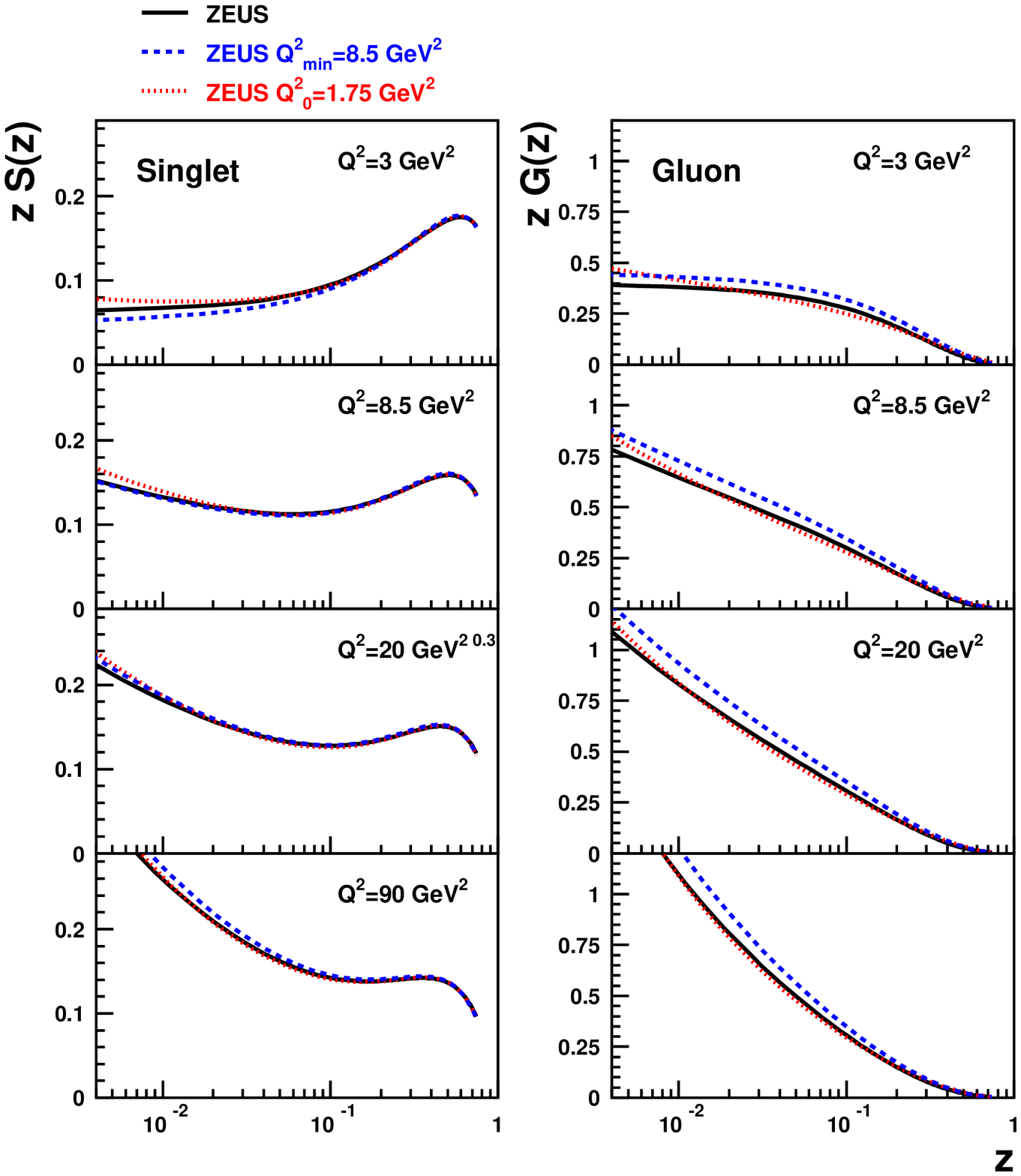}
\end{center}
\caption{ Singlet and gluon distributions 
of the Pomeron as a function of $z$ derived from  QCD fits on ZEUS data. 
Results are presented with $Q^2_{min}=4.5$~GeV$^2$ (full lines),
$Q^2_{min}=8.5$~GeV$^2$ (dashed lines). Distributions obtained from a
fit with  $Q^2_0=1.75$~GeV$^2$ (dotted lines) is also shown.
}
\label{fig3}
\end{figure}

\begin{figure}[htbp]
\begin{center}
\includegraphics[totalheight=18cm]{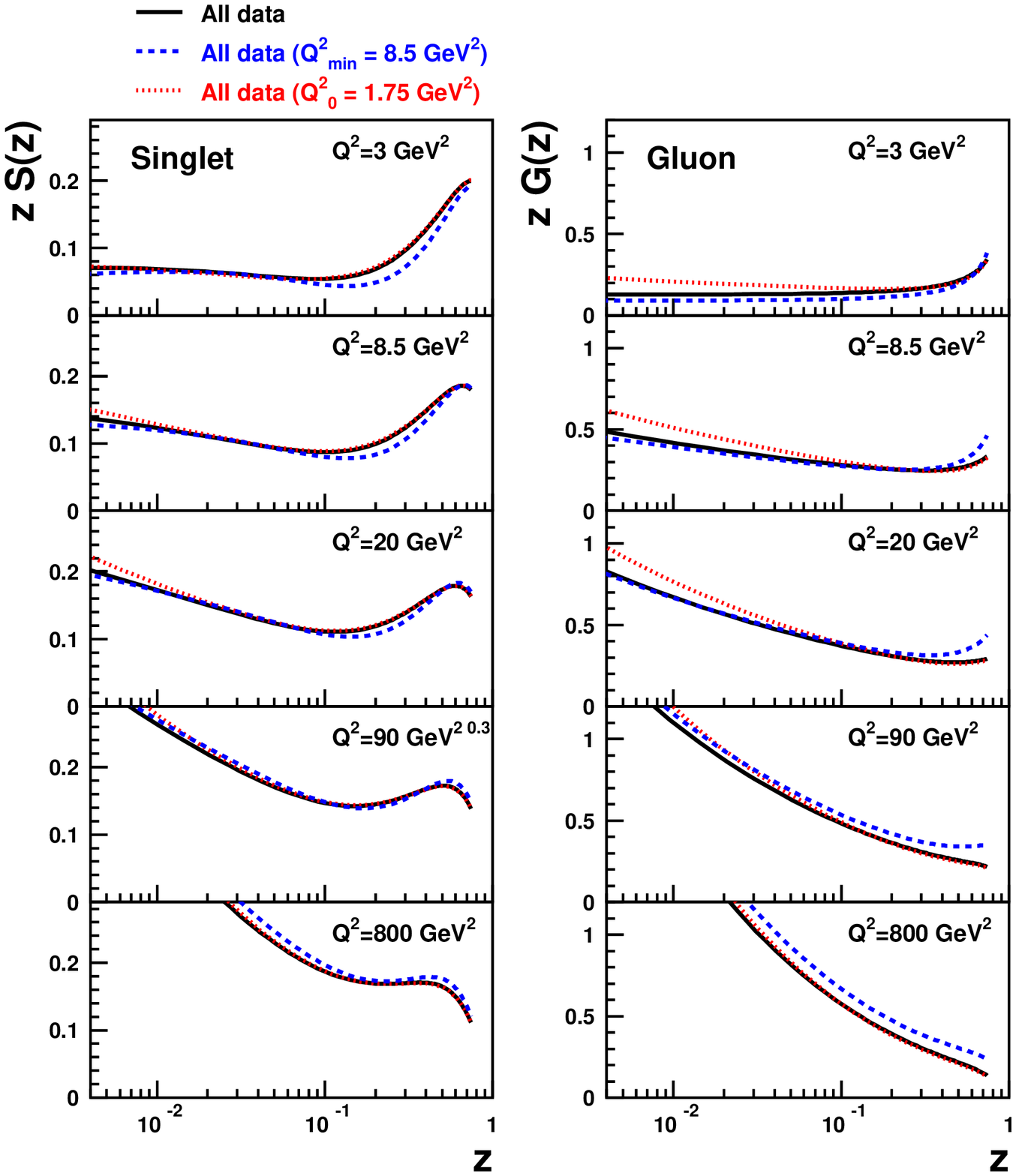}
\end{center}
\caption{ Singlet and gluon distributions 
of the Pomeron as a function of $z$ derived from  QCD fits on all H1 and ZEUS data sets. 
Results are presented with $Q^2_{min}=4.5$~GeV$^2$ (full lines),
$Q^2_{min}=8.5$~GeV$^2$ (dashed lines). Distributions obtained from a
fit with  $Q^2_0=1.75$~GeV$^2$ (dotted lines) are also shown.
}
\label{figall1}
\end{figure}

\begin{figure}[htbp]
\begin{center}
\includegraphics[totalheight=18cm]{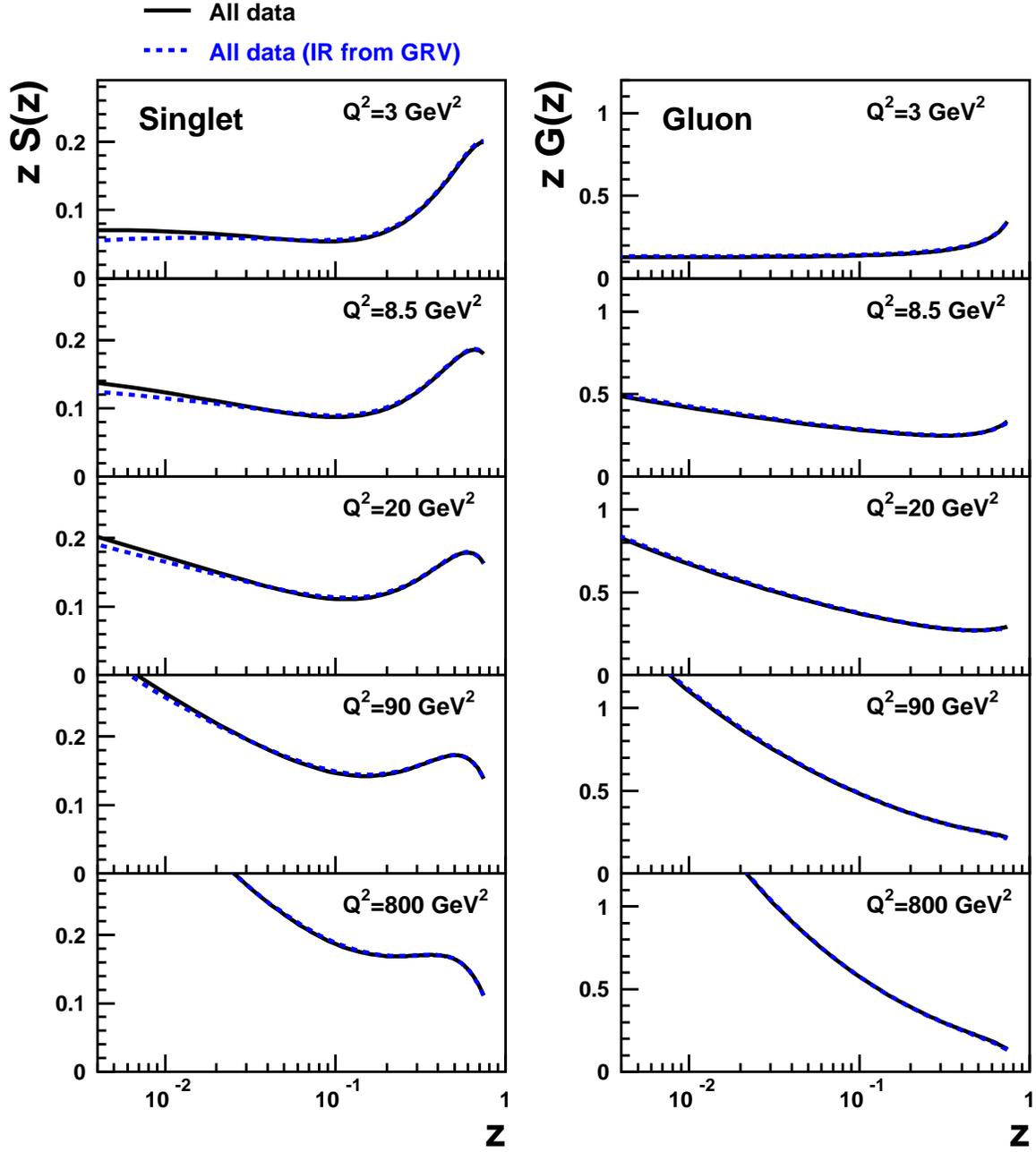}
\end{center}
\caption{ Singlet and gluon distributions 
of the Pomeron as a function of $z$ derived from  QCD fits on all H1 and ZEUS data sets. 
Results are presented for a QCD fit using the 
Owens (full lines) or GRV (dashed lines) parametrisations of the Reggeon contribution. 
}
\label{figall2}
\end{figure}

\begin{figure}[htbp]
\begin{center}
\includegraphics[totalheight=18cm]{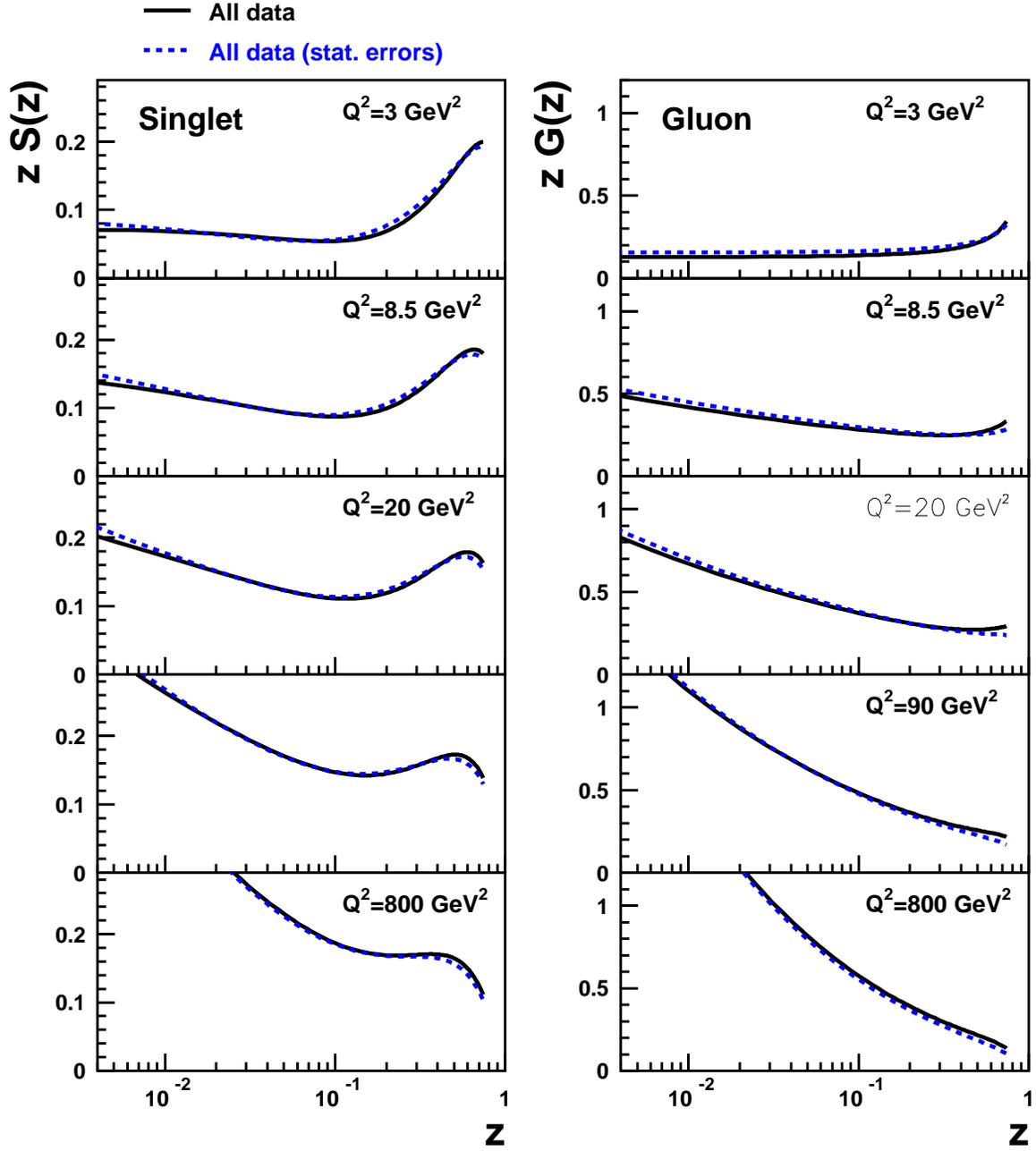}
\end{center}
\caption{ Singlet and gluon distributions
of the Pomeron as a function of $z$ derived from  QCD fits on all H1 and ZEUS data sets.
Results are presented for fits using the
total errors (full lines) or the statistical errors (dashed lines).
}
\label{figall3}
\end{figure}

In this appendix, we want to give the systematics checks we performed on the DGLAP based
QCD fits to the H1 rapidity gap data alone (H1RAP), ZEUS $M_X$
data (ZEUSMX) or the combined data sets:
\begin{itemize}
\item We check the dependence of the DPDFs on variations of the starting scale $Q_0^2$ in Fig. \ref{fig0}
for the H1 data,
very small changes are observed while changing the starting scale form
3 to 1.75 GeV$^2$. Note that the results on the ZEUS $M_X$ data and on the
combined data sets are given in Figs. \ref{fig3} and \ref{figall1} and lead to the same
conclusion
\item We check the fit stability by changing the cut on $Q_{min}^2$, the lowest value
of $Q^2$ of data to be included in the fit. The results are given in Figs.
\ref{fig1} (H1RAP), \ref{fig3} (ZEUSMX) and 
\ref{figall1} for the combined data set where we show the results of the fits to the H1 and
ZEUS data after applying a cut on $Q^2_{min}$ of 4.5, 8.5 and 12 GeV$^2$.
Differences are noticeable at small $\beta$ but well within the fit
uncertainties. No systematic behaviour is observed within $Q^2_{min}$ variations.
\item We check the dependence on the pion structure function by changing it 
from the Duke Owens parametrisation to the GRV one \cite{grv}. The results are
given in Fig. \ref{figall2} for the fits of all data sets
and only very small differences are observed between
both fits showing the weak dependence of the fit on the precise knowledge of the
Reggeon structure function
\item We also check the fit stability by doing it with statistical errors only
or with statistical and systematics errors added in quadrature. The results are
shown in Fig. \ref{figall3} and show very small different on the DPDFs between both fits.
\end{itemize}


\newpage




\begin{thebibliography}{99}


\bibitem{f2d97}
  H1 Collab., A.~Aktas { et al.},
  arXiv:hep-ex/0606004.  
\bibitem{f2d97b}
  H1 Collab., A.~Aktas { et al.},
  arXiv:hep-ex/0606003 .  
\bibitem{zeus}
  ZEUS Collab., S.~Chekanov { et al.},
  Nucl.\ Phys.\ {\bf B713} (2005) 3
  [arXiv:hep-ex/0501060].
\bibitem{zeusb}
  ZEUS Collab., S.~Chekanov {\it et al.},
  Eur.\ Phys.\ J.\ {\bf  C38} (2004) 43
  [arXiv:hep-ex/0408009].




\bibitem{ingelman} G. Ingelman, P. Schlein,
{\it Phys. Lett.} {\bf  B 152} (1985) 256.

\bibitem{soper} J.Collins, {\ Phys.Rev.} {\bf D57} (1998) 3051;
Erratum-ibid. {\bf D61} (2000) 019902.

\bibitem{dglap} G.Altarelli and G.Parisi,
{\ Nucl. Phys.} {\bf B126}  18C (1977) 298.
V.N.Gribov and L.N.Lipatov, { Sov. Journ. Nucl. Phys.} (1972) 438 and 675.
Yu.L.Dokshitzer, { Sov. Phys. JETP.} {\bf 46} (1977) 641.

\bibitem{jung}
H. Jung {\it Comput. Phys. Commun.} {\bf 86} (1995) 147.

\bibitem{pap2006}  
  C.~Royon, L.~Schoeffel, R.~Peschanski and E.~Sauvan,
  Nucl.\ Phys.\ B {\bf 746} (2006) 15
  [arXiv:hep-ph/0602228].

\bibitem{grv} 
  M. Gl\"uck, E. Reya, A. Vogt, {t Z. Phys.} {\bf C53} (1992) 651.

\bibitem{owens} 
  J.F. Owens, {\ Phys. Rev. } {\bf D30} (1984) 943.

\bibitem{qcdnum}
  M.~Botje,
  Eur.\ Phys.\ J.\ {\bf C14} (2000) 285
  .
\bibitem{lolo} 
  L. Schoeffel, {\ Nucl.Instrum.Meth.} {\bf A423} (1999) 439.


\bibitem{ghrgrs} 
  M.~Gl\"uck, E.~Hoffmann, E.~Reya, {\ Z.~Phys.} {\bf  C13} (1982) 119; \\
  M.~Gl\"uck, E.~Reya, M.~Stratmann, {\ Nucl.~Phys.} {\bf  B422} (1994) 37.


\bibitem{diehl}
  M.~Arneodo and M.~Diehl,
  arXiv:hep-ph/0511047.


\bibitem{bartels} J.Bartels, J.Ellis, H.Kowalski, M.Wuesthoff,  
Eur.Phys.J. {\bf C7} (1999) 443;
  J.~Bartels and C.~Royon,
  Mod.\ Phys.\ Lett.\  {\bf A14} (1999) 1583.

\bibitem{thrust}H1 Collab., C.Adloff et al., Eur. Phys. J.{\bf  C1} (1998) 495.

\bibitem{hfs} H1 Collab., C.\ Adloff et al., Phys. Lett. {\bf B428} (1998) 206-220;  \\
              H1 Collab., C.\ Adloff et al., Eur. Phys. J. {\bf C5} (1998) 3, 439.


\bibitem{dipole} A.H.Mueller and B.Patel, 
Nucl. Phys. {\bf B425} (1994) 471; \\
A.H.Mueller, Nucl. Phys. {\bf B437} (1995) 107; \\
A.H.Mueller, Nucl. Phys. {\bf B415} (1994) 373; \\
H.Navelet, R.Peschanski, Ch.Royon, S.Wallon,
Phys. Lett. {\bf B385} (1996) 357.

\bibitem{bfkl} V.S.Fadin, E.A.Kuraev, L.N.Lipatov
Phys. Lett. {\bf B60} (1975) 50. \\ 
I.I.Balitsky, L.N.Lipatov, Sov. J. Nucl. Phys. 28 (1978) 822.

\bibitem{robi} A.Bialas, R.Peschanski, C.Royon, Phys. Rev. {\bf D57} (1998) 6899; \\
S.Munier, R.Peschanski, C.Royon, Nucl. Phys. {\bf B534 } (1998) 297.

\bibitem{gbw1}
  K.~Golec-Biernat and M.~Wusthoff,
  { Phys.\ Rev.}\ {\bf D59} (1999) 014017.
\bibitem{gbw2} 
  K.~Golec-Biernat and M.~Wusthoff,
  { Phys.\ Rev.}\ {\bf D60} (1999) 114023.
\bibitem{Bartels:2002cj}
  J.~Bartels, K.~Golec-Biernat and H.~Kowalski,
  { Phys.\ Rev.}\ {\bf D66} (2002) 014001.
\bibitem{Golec-Biernat:2006ba}
  K.~Golec-Biernat and S.~Sapeta, Phys.\ Rev.\ {\bf D74} (2006) 054032.

\end{thebibliography}
\end{document}